\documentclass[document,screen,review=false]{acmart}

\AtBeginDocument{%
  \providecommand\BibTeX{{%
    \normalfont B\kern-0.5em{\scshape i\kern-0.25em b}\kern-0.8em\TeX}}}

\setcopyright{acmcopyright}
\copyrightyear{2018}
\acmYear{2018}
\acmDOI{XXXXXXX.XXXXXXX}

\acmJournal{JACM}
\acmVolume{37}
\acmNumber{4}
\acmArticle{111}
\acmMonth{8}

\acmPrice{15.00}
\acmISBN{978-1-4503-XXXX-X/18/06}

\begin{document}

\title{Investigating Participation Mechanisms in EU Code Week}

\author{Christel Sirocchi}
 
\email{c.sirocchi2@campus.uniurb.it}
\orcid{0000-0002-5011-3068}
 
\affiliation{
  \institution{DiSPeA, University of Urbino}
  \streetaddress{Piazza della Repubblica, 13}
  \city{Urbino}
  \country{Italy}
  \postcode{61029}
}

\author{Annika Ostergren Pofantis}
\affiliation{
  \institution{DG Connect, European Commission}
  \streetaddress{}
  \city{Brussels}
  \country{Belgium}
  \postcode{}
}
\email{Annika.OSTERGREN@ec.europa.eu}

\author{Alessandro Bogliolo}
\affiliation{
  \institution{DiSPeA, University of Urbino}
  \streetaddress{Piazza della Repubblica, 13}
  \city{Urbino}
  \country{Italy}
  \postcode{61029}
}
\email{alessandro.bogliolo@uniurb.it}

\renewcommand{\shortauthors}{Sirocchi et al.}

\begin{abstract}
Digital competence (DC) is a broad set of skills, attitudes, and knowledge for confident, critical and responsible use of digital technologies in every aspect of life. DC is fundamental to all people in conducting a productive and fulfilling life in an increasingly digital world. However, prejudices, misconceptions, and lack of awareness reduce the diffusion of DC, hindering digital transformation and preventing countries and people from realising their full potential. Teaching Informatics in the curriculum is increasingly supported by the institutions but faces serious challenges, such as teacher upskilling and support, and will require several years to observe sizeable outcomes. In response, grassroots movements promoting computing literacy in an informal setting have grown, including EU Code Week, whose vision is to develop computing skills while promoting diversity and raising awareness of the importance of digital skills. Code Week participation is a form of public engagement that could be affected by socio-economic and demographic factors, as any other form of participation. The aim of the manuscript is twofold: first, to offer a detailed and comprehensive statistical description of Code Week's participation in the EU Member States in terms of penetration, retention, demographic composition, and spatial distribution in order to inform more effective awareness-raising campaigns; second, to investigate the impact of socio-economic factors on Code Week involvement. The study identifies a strong negative correlation between participation and income at different geographical scales. It also suggests underlying mechanisms driving participation that are coherent with the "psychosocial" and the "resource" views, i.e. the two most widely accepted explanations of the effect of income on public engagement.
\end{abstract}



\maketitle

\section{Introduction}

In a highly interconnected world, where social cohesion and sustainable development heavily depend on the competence of all citizens, recognising and promoting valuable skills in the population is of critical importance. In recent years, there has been a growing understanding that basic literacy must include digital competence, which goes beyond basic digital skills but wholly contributes to a young person's skill set and ability to conduct a productive and fulfilling life in an increasingly digital world. Digital competences have emerged as fundamental for all citizens, particularly in the light of the COVID-19 pandemic \cite{blasko2021learning}.

Late, insufficient or inadequate introduction of Informatics in compulsory education hinders the widespread diffusion of digital literacy \cite{bocconi2022reviewing}, prevents citizens from recognising that Computer Science is the fundamental discipline underpinning the digital transformation \cite{caspersen2018informatics}, and leads to misunderstanding of common phrases, such as Digital Competence (DC) or Computational Thinking (CT). It is still a common belief that computing skills are professional skills to be acquired in the pursuit of a specific career path and of little use in everyday life. On the contrary, DC is a set of essential skills, attitudes and knowledge, broadly defined as the confident, critical and responsible use of digital technologies for working, learning, leisure, self-development and participating in society \cite{councilrecomm} \cite{carretero2017digcomp}. DC entails understanding the underlying logic and mechanisms, the opportunities, risks and limitations of evolving digital technologies, as well as the ethical and legal principles involved when engaging with them. It empowers citizens to use modern technologies in supporting collaboration, creativity and innovation toward personal, social, and commercial goals while enabling active citizenship and social inclusion \cite{esteve2020digital}. The safe and responsible use of digital tools requires a critical, reflective, open-minded and forward-looking attitude to these technologies\cite{councilrecomm}. Hence, digital transformation, other than acquiring digital tools, entails a deeper awareness and a change of attitude toward the role of technology in our society. CT is one of the most critical areas making up this digital competence \cite{esteve2020digital} and refers to the thinking required in formulating a problem and expressing its solution so that an executor, be it human or machine, can carry out \cite{wing2017computational}. CT entails decomposing problems, recognising patterns, extracting basic principles, developing algorithms, implementing automated solutions, and debugging. \\

Critical aspects to building advanced digital skills are providing lifelong learning opportunities, starting as early as possible (primary education), embedding digital skills within every subject, and regularly revisiting skills and approaches depending on the ever-evolving technologies \cite{ala2008digital}. Early exposure to computing concepts is also relevant in the context of inclusion, diversity and gender, promoting self-efficacy and tackling stereotypes before prevailing norms become entrenched. Schools, particularly at the primary and lower-secondary levels, are at the forefront of this digital transformation of societies and economies, being the first providers of digital education for the future generation of citizens. Young students begin to create with digital technology through innovative and encouraging approaches and gain a solid understanding of the digital world while developing creativity, problem-solving, communication, collaboration and analytical skills \cite{manches2017computing}. The EU Member States and many more countries worldwide have increasingly supported teaching computational thinking and other computer science notions in a formal education setting \cite{bocconi2022reviewing}. Integrating these concepts into the primary and lower secondary curriculum is relatively new and faces several challenges, the most significant of which is teacher upskilling and support. Recommendations for the effective integration of DC and CT skills into compulsory education focus on providing teacher professional development, raising awareness about the purpose and benefits of CT skills, and prioritising measures to address gender balance, equity, and inclusion \cite{bocconi2022reviewing}. Among effective support measures for teachers are collaborative support among teachers, such as sharing experiences and networking, and ample access to high-quality learning and teaching materials.\\


Recognising the urgent need to provide the young population with accessible opportunities to develop digital skills, many institutions and organisations have promoted coding literacy in a non-formal context. Among them is EU Code Week \cite{CodeWeek}, launched in 2013 and supported by the European Commission as part of the Digital Single Market Strategy. Code Week aims to help more and more people learn the fundamentals of coding and computational thinking while raising awareness about the importance of computing skills for all citizens, promoting diversity and battling stereotypes. It is a further opportunity for young students to explore digital creativity and coding and develop digital skills and attitudes. It also provides schools and teachers with free professional development and offers a platform where teachers can connect and share experiences. Teachers voluntarily join the initiative by organising coding activities in their institute and adding them to the codeweek.eu map. The level of Code Week participation, in terms of events, teachers and institutes involved, is the most suitable metric to assess the initiative's impact. Code Week is an awareness-raising campaign rather than a coding education provider, so its success is better defined by how many and which people it reached. Characterising Code Week participation patterns would help assess the campaign's effectiveness and help design better outreach strategies. Like other forms of voluntary participation, teacher engagement is expected to be affected by the contextual factors of the socio-economic and educational setting \cite{filetti2018income}. Investigating factors both at the society and individual level would shed light on the underlying mechanisms driving Code Week participation.\\



The aim of this paper is twofold: to provide a detailed and comprehensive statistical description of the participation in EU Code Week and to analyse its relationship with socio-economic factors by integrating Code Week data with multiple open data sets. Code Week participation is intended as that of teachers and hosting institutions organising events within the initiative rather than that of attendees taking part in the activities. While the first form of engagement is primarily voluntary, the latter is not because most events take place in schools and students are not given a choice to participate. Four types of participation metrics, i.e. penetration, demographic, spatial and retention metrics, were calculated for EU countries to characterise engagement at different levels and distinguish between endogenous factors, i.e. affected by the country's internal dynamics, and exogenous factors, i.e. physiological to the Code Week campaign format. The event number distribution, the growth trends, and the location of new teachers indicate a degree of spatial associativity: teachers are more likely to participate if there are other participating teachers in their proximity. This suggests the presence of an offline network fostering collaboration among teachers in the territory and fostering Code Week participation.\\


The study also identified a strong negative linear correlation between Code Week participation, quantified as event number normalised per population, and the GDP per capita. This relationship was further investigated in Italy at the regional, provincial and municipal levels. Other socio-economic metrics related to the development level, wealth inequality, and infrastructure quality were considered but did not have significant explanatory power when controlling for GDP, suggesting that the country's resources directly affect participation. Previous results on the effect of income and income inequality on social and civic engagement identified a positive correlation. The study argues that the peculiar characteristics of Code Week are responsible for the unusual trend observed and discuss how the case under consideration is in agreement with both the "psychosocial" and "resource" explanations of the effect of income on participation. Finally, it was observed that the most participating less developed regions are also those with lower levels of digital literacy and access to technology, and less established computer education in their curriculum, meaning that Code Week is more prevalent where it is most needed.\\

The remainder of the paper is structured as follows: Section 2 revises relevant literature related to the topic, Section 3 describes the data sets used in the investigation, Section 4 defines the participation metrics calculated on the Code Week data, Sections 5 and 6 present and discuss the analysis results, respectively, Section 7 draws conclusions and discusses future work.

\section{Previous Work}

\subsection{Computing Education in Europe}

Computational thinking (CT) is generally intended as the ability to formulate a problem and express its solution(s) such that a human or machine executor can effectively carry out \cite{wing2017computational}. CT and other digital skills have been recognised as fundamental in enabling citizens to understand and operate in the digital society. The European Commission elected "A Europe fit for the digital age" as one of the six priorities for 2019 - 2024, intending to empower the population with a new generation of technologies \cite{sixpriorities}. The Digital Education Action Plan 2021-2027 includes "Enhancing digital skills and competences for the digital transformation" in its priority areas \cite{digitaleducation}. Ministries of Education of most EU Member States have responded well to Europe's call to action, revising their statutory curriculum by introducing computing education in primary and secondary schools.\\

The Joint Research Centre (JRC) of the European Commission has produced a technical report in 2016 \cite{bocconi2016developing} and an update in 2022 \cite{bocconi2022reviewing} which present recent policies and grassroots movements in promoting computer education for schoolchildren. The study reports that 18 EU Member States (AT, BE, CY, EL, ES, FI, FR, HR, HU, IE, LT, LU, MT, PL, PT, RO, SE, SK) and 5 other EU countries (CH, NO, RS, RU, UK-ENG) have introduced CT and basic computing concepts in their statutory curriculum. In these countries, except for 6 (BE, CY, ES, HR, IE, RO), basic computer science concepts are compulsory in both primary and lower secondary schools. Other EU Member states have a strategic plan to introduce CT skills (CZ, IT, SI), are running school pilots (DK), or are finalising a draft curriculum (BE) \cite{bocconi2022reviewing}.\\

CT skills integration follows three main approaches \cite{bocconi2016developing}: (i) as a cross-curricular theme addressed in all subjects, with all teachers sharing the responsibility for developing CT skills; (ii) as part of a separate subject related to computing (e.g. Informatics); (iii) integrated within some other curriculum subject (e.g., Maths, Technology). The cross-curricular approach is the most common in primary education, often combined with one of the other two. At this level, teachers cover several subjects and often do not have specialised subject-area training and expertise. On the other hand, in lower secondary education, CT skills are either addressed in a separate subject or within other subjects. In this case, appointed teachers generally have domain-specific expertise \cite{bocconi2022reviewing}.\\

In all assessed countries, teachers have a degree of autonomy in delivering the curriculum and handling implementation. Teachers adopt age-appropriate pedagogical approaches, often borrowed from other learning contexts. Primary teachers report using playful, hands-on activities with block-based visual programming, programmable robots and unplugged methods, where students are encouraged to work in groups and learn from their mistakes. Lower-secondary teachers focus on fostering problem-solving skills and promoting student autonomy and self-efficacy by working on real-life problems and encouraging students to create their programs, applications, video games, etc. A key aspect spanning both levels is the iterative approach and debugging to identify and correct errors while emphasising the value of making mistakes in strengthening one's expertise \cite{bocconi2022reviewing}.\\

The scarcity of teaching tools and resources and the shortage of teachers with suitable background and adequate capabilities are among the main challenges to the successful integration of CT skills into the curriculum \cite{li2021computational}: Computer Science teachers are scarce and difficult to recruit \cite{li2021computational}; generalist teachers often hesitate to include CT in their practices due to the lack of confidence in their programming capabilities and deep understanding of CT concepts \cite{zhang2020progression}. To overcome the lack of resources and competences, sustained financial support should be provided to schools so that teachers can attend professional development programs and adequate digital infrastructure is built. Moreover, whereas CT is taught as a cross-curricular theme, collaboration and division of tasks and responsibilities among teachers should be carefully planned and coordinated \cite{bocconi2016developing}.\\

\subsection{Computing Education in informal settings}

Many institutions and organisations have offered opportunities to develop CT and digital skills in an informal environment to support and supplement computing education in the curriculum. Local institutions frequently organise workshops and summer camps to introduce students to programming and emphasise possible careers in computer science. In the last decades, many international initiatives have stemmed, most notably Code.org, CoderDojo and EU Code Week. Code.org is a non-profit organisation that facilitates access to computing education for students, particularly for minorities and other disadvantaged groups. It is a leading provider of K-12 computer science curricula in the United States and is behind the successful "Hour of Code" campaign, which has engaged over 15\% of students worldwide \cite{codeorg}. The CoderDojo movement is a global network of freely accessible programming clubs managed by local volunteers. Young students begin learning programming language through hands-on projects, e.g. creating an app or a game, building a website, and exploring digital technologies \cite{coderdojo}. \\

EU Code Week is a grassroots initiative promoted by the European Commission to bring coding and digital literacy to everybody. It relies on a network of volunteer ambassadors in participating countries, who promote the initiative, help spread the vision of Code Week, and encourage the organisation of coding events within their area of relevance. Leading teachers, forming a network of over 450 teachers across Europe, connect educators, companies and communities interested in organising activities. They also hold professional development training and promote the initiative locally, acting as a reference point in the community \cite{CodeWeek}. EU Code Week also offers a large body of teaching and learning materials in the form of MOOC, presentations and tool-kits to assist teachers in organising events. Code Week events include not only in-school activities but also introductory and specialised workshops, motivational talks, and open door days \cite{moreno2015europe}.\\

\subsection{Effect of socio-economic factors on participation}

Civic and social participation are the two prominent ways individuals engage with their community. The former generally refers to membership and involvement in voluntary associations, such as neighbourhood committees, charitable organisations, political parties, peace or environmental movements. The latter is often defined as the informal bonds between people, particularly friends and family, and differs from the first one for the lack of formal and institutionalised ties. Both forms of participation have crucial effects on all aspects of society. Civic engagement in general, and political participation in particular, are fundamental to democracy \cite{verba1995voice}. It is the mechanism by which citizens are able to communicate their needs and mobilise pressure for a government response \cite{dubrow2014political}. One of its cornerstones is that participation must be voluntary so that citizens can express their humanitarian and civic concerns and pursue their self-interest. Voluntarism helps shape the allocation of economic, social and cultural benefits, contributing to achieving collective goals \cite{verba1995voice}. Hence, voluntarism is fundamental to democracy because it reflects the individual's autonomy within the society. Social participation contributes to the so-called "social capital", defined as the set of social features (networks, norms, trust) that enable participants to work more effectively together and pursue shared objectives \cite{putnam1995tuning}. The social capital theory attests that social networks have value and that individual virtue is most valuable when embedded in a web of social relations. Social participation affects many individual outcomes, such as health, employment and education, through the beneficial effects of this shared social capital. It also has a crucial impact on the economy through sharing knowledge and personal experience, increasing society's human capital and promoting social trust, which is known to improve the functioning of markets \cite{lopez1997trust}. \\ 


Many research studies primarily focused on individual characteristics affecting the choice to engage or not in social activities. However, several empirical studies have shown how features of societies or communities as a whole affect the degree of participation of their members. Foremost, levels of inequality in society have been shown to affect social and civic engagement. Oliver \cite{oliver1999effects} first suggested that the economic context influences civic participation in two ways: wealthy cities have fewer social needs promoting political action, while heterogeneous cities have more competition for public goods prompting citizens' interest and participation. Hence, civic participation is lowest in homogeneous, affluent cities and highest in diverse, middle-income cities. Solt \cite{solt2008economic} demonstrated that higher levels of income inequality powerfully depress political interest, particularly in terms of the frequency of political discussion and participation in elections. Filetti et al. \cite{filetti2018income} confirmed these results and observed that economic contraction influence the within-country distribution of active citizens but not the overall level of participation. Waldman \cite{levin2013income} investigated the link between income, personal autonomy and civic engagement, claiming that the lack of economic resources limits the individual's self-sufficiency, which in turn hinders participation. Having confirmed that higher household income leads to higher civic engagement, the author speculated that wage policies aimed at reducing income inequality would result in greater democracy.\\

Alesina et al.\cite{alesina2000participation} showed that participation in a variety of social activities, e.g. religious groups, youth groups and sports clubs, is significantly lower in unequal and more ethnically or racially segmented societies. Ulsaner et al. \cite{uslaner2005inequality} argued that the effect of inequality on participation arises when disadvantaged people do not participate, either because they do not dispose of the necessary resources or because they have lost trust in society and believe participation would be fruitless. The authors refer to income inequality as "the strongest determinant of trust" over time and geographical areas. Wilkinson et al. \cite{wilkinson2010spirit} observed that higher levels of inequality relate to lower levels of social trust, poor health, higher crime rates, and lower levels of participation. They suggested that inequality affects these outcomes through psychosocial processes of status differences and related distress caused by their disadvantaged position. Rotolo et al. \cite{rotolo2014social} confirmed that income inequality correlates with all types of voluntary activity, while the effect of racial heterogeneity and segregation vary with volunteering type. Collins et al. \cite{collins2018effect} found that lower civic engagement is associated with feeling less safe but also pointed out that social capital, rather than civic engagement, mediates the relationship between inequality and the sense of safety. Studies on the effect of income inequality on civic participation intended as charitable giving yielded contradicting results. According to Payne et al. \cite{payne2015does}, increases in inequality increase donations, although results are sensitive to the geographical dispersion of high- and low-income households. Duquette \cite{duquette2018inequality}, on the other hand, identified an inverse correlation between donations by high-income households and inequality in the USA, confirming this result in an experimental setting \cite{duquette2021inequality}, suggesting that charitable giving strengthens inequality over time.\\


Although previous studies lead to different conclusions on the socio-economic aspects that affect civic and social participation, depending on the context and on the variables they focus on, most of them indicate that social inequality and low income hinder participation.
In particular, the proposed explanations for inequality's impact on participation fall under two main categories. The "psychosocial explanation" claims that inequality undermines social cohesion, eroding social capital, trust, safety and security of citizens \cite{elgar2011income} \cite{collins2018effect}. Inequality - in terms of income, wealth or social class - inhibits volunteering, club membership and charitable giving because it removes the conditions promoting social interaction and collaboration, including opportunities for personal acquaintance, equal status between citizens, and the ability to share common goals within the society \cite{allport1954nature}. Homophily, i.e. the tendency of individuals to associate with others similar to them, also contributes to lowering trust, solidarity, and as a result, participation \cite{mcpherson2001birds}. The "resource explanation" argues instead that the availability of resources affects participation because resources are needed to achieve desirable outcomes. In this scenario, equal societies participate more because they can provide collective goods more and more equally to all community members. In contrast, in unequal societies, low income and systematic underinvestment across a wide range of human, physical, health, and social infrastructure limit resources held by individuals and increase the unmet need for public goods \cite{lynch2000income}. The two explanations are not mutually exclusive. Several studies have investigated the tenability of both perspectives to determine whether inequality affects outcomes through psychosocial processes or through individual or collective resources that are made available to individuals, \cite{elgar2011income} \cite{lancee2012income}. For instance, Lancee et al. \cite{lancee2012income} investigated both perspectives and found that the main effects of income inequality on civic and social participation manifest via resources held at the individual and societal levels.\\ 

A recent study reviewed 70 studies addressing the effect of various forms of inequality on civic engagement in terms of charitable giving, non-profit membership and volunteering and found that higher inequality is most often negatively correlated with civic engagement \cite{schroder2021socio}. However, the study also pointed out that the generalisability and empirical understanding obtained from the considered studies are limited. For all forms of participation, and volunteering in particular, several studies reported an opposite trend, albeit often with a non-significant coefficient. Most civic engagement research (60-80\%) is from the USA, characterised by a peculiar profile of social inequality, and is rarely coherent with results from other countries\cite{schroder2021socio}. Also, cross-country and within-country data yield mixed results, suggesting that different mechanisms operate a different geographical scales, even though most studies only address one geographical scale, guided by data availability. Several confounders, namely individual income and education, appear to moderate the effect of inequality on engagement, but findings are mixed, hence inconclusive\cite{schroder2021socio}. Finally, the study remarks that the three chosen forms of civic engagement are not an entirely homogeneous set of activities, so different mechanisms and motivations can be at work, resulting in different participation patterns. The effect of socio-economic factors on civic engagement appears to largely depend on the type of activity considered, the context, and the geographic scale, suggesting that previous studies conducted in different settings offer little to no explanatory power to the case under investigation.


\section{Data description}

\subsection{Code Week}
The anonymised dataset of Code Week, accessed in January 2022, comprises 321,213 events created within the Code Week initiative starting from the 2014 edition. A description of the variables included in the dataset is reported in Table \ref{tab:CodeWeek}. Only activities between 2014 and 2021 were considered, thus excluding 834 events programmed to take place in 2022. The geographic coordinates of the events were used for remapping events on EU territory according to the Nomenclature of territorial units for statistics (NUTS) \cite{Eurostat:NUTS}. The current NUTS 2021 classification was applied, valid from 1st January 2021, and dividing the economic territory of the EU and the UK into 92 macro areas at NUTS 1, 242 regions at NUTS 2 and 1,166 provinces at NUTS 3 level. 21,419 events mapped outside of this territory were excluded from the analysis. Geographic coordinates approximated to 3 decimal places ( $\approx$ 100 meters) identify a hosting location.\\

\begin{figure}[h]
    \centering
    \includegraphics[width=\linewidth]{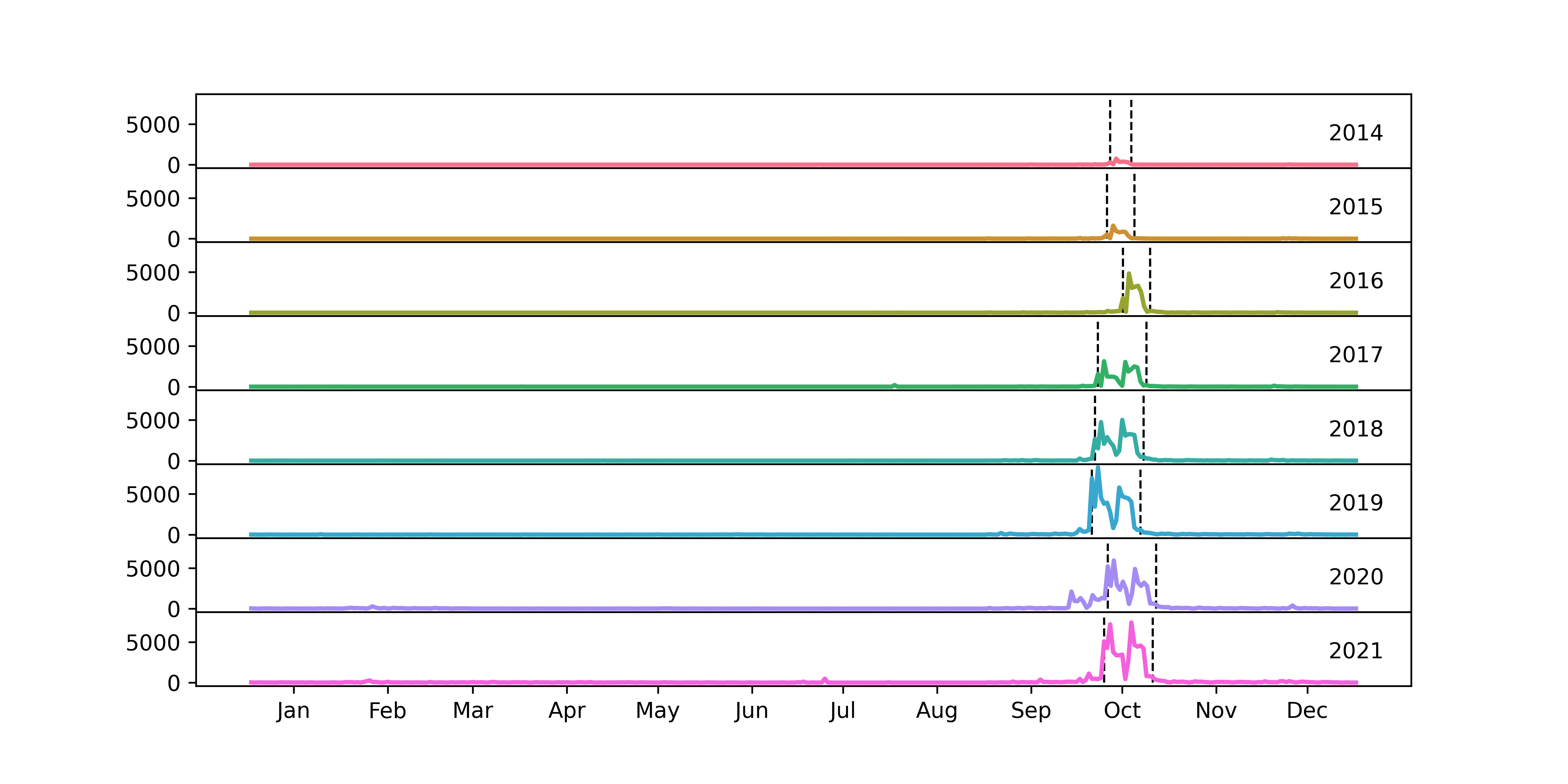}
    \caption{Distribution of EU events from 2014 and 2022. Code Week official week(s) are delimited by vertical lines}
    \Description{}
    \label{fig:events_in_time}
\end{figure}

Code Week activities can be organised and added to the database all year round. However, the vast majority of the events occur during the official Code Week period (covering 2 weeks since 2017 edition) or the days preceding it, as shown in Figure \ref{fig:events_in_time}. In Austria, however, only 13\% of events are scheduled during the official Code Week period and 25\% of all events in the country (607 out of 2442) were registered to take place on two dates in 2021, suggesting a bulk upload mechanism that might facilitate creating and uploading events, hence overestimating this country's participation. Events with more than 1,000 participants (632 events) or with an average age below 3 (455) or above 60 (252) were regarded as potentially erroneous entries and excluded from the analysis.

\begin{table}
    \caption{Code Week events dataset}
    \label{tab:CodeWeek}
    \begin{tabular}{p{0.2\textwidth}p{0.7\textwidth}p{0.1\textwidth}}
    \toprule
    Variable Name & Variable Description & Availability\\
    \midrule
     geoposition & geographical coordinates of the event location in reference system "EPSG:4326" & All\\
     country\_iso & country where the event is taking place, as declared by the event creator & All\\
     start\_date & starting time and date of the event & All\\
     end\_date & ending time and date of the event & All \\
     creator\_id & ID number uniquely identifying an event creator & All\\
     participants\_count & number of participants at the event & Report\\
     average\_participant\_age & average age of participants to the event & Report\\
     percentage\_of\_females & percentage of female participants & Report\\
     reported\_at & time and date at which the report was submitted by the event creator & Report\\
     codeweek\_for\_all & ID code characterising event within the CodeWeek4all initiative & CodeWeek4All\\
     organizer\_type & type of organising institute, selecting 1 from: school, non profit, private business, other, library, club & Report  \\
     activity\_type & type of organised event, selecting 1 from: open-online, invite-in-person, open-in-person, invite-online, open-online \& open-in-person & All\\
     language & language in which the event is conducted & \\
     tags & 1 or more tag words chosen by the event creator & All\\
     themes & the main theme of the event, selecting at least 1 from from: Basic programming concepts, Playful coding activities, Motivation and awareness raising, Visual/Block programming, Software development, Promoting diversity, Robotics, Mobile app development, Art and creativity, Unplugged activities, Web development, Other, Hardware, Augmented reality, Data manipulation and visualisation, Internet of things and wearable computing, Game design,
     3D printing, Artificial intelligence & All\\
     audiences & the target audience for the event, selecting at least 1 from: pre-school children, elementary school students, high school students, graduate students, employed adults, unemployed adults & All\\
   \bottomrule
\end{tabular}
\end{table}

\subsection{Socio-economic and demographic statistics}

Population data relating to NUTS 0, NUTS 2 and NUTS 3 statistical territories, i.e. European countries, regions and provinces, respectively, were retrieved from the Regional demographic statistics page of the Rural Development Data Section published by Eurostat \cite{Eurostat:NUTS_population}. The dataset was last updated on 03-06-2021 and includes census data from 2016 until 2020, where each value refers to the population size on 1st January of the considered year. The study uses 2016 census data to include the United Kingdom in the analysis, as more recent data is unavailable for this country due to its departure from the European Union. Gross Domestic Product (GDP) per capita for EU countries was obtained from the "Decent work and economic growth" section of the Sustainable development indicators \cite{Eurostat:GDP_sdi} in Tables of EU policy published by Eurostat. GDP reflects the total value of goods and services produced by an economy within the year minus the value of those used in production. GDP per capita is calculated as the GDP divided by the average population of a given year, allowing to compare economies significantly different in absolute size. As a measure of economic activity, it is often used to proxy a country's material living standards. However, it is not a comprehensive measure of economic welfare because, for instance, it does not account for unpaid household work or the harmful effects of economic activity (e.g. environmental damage caused by industrialisation). The dataset contains statistics from 2000 to 2021 and was last updated on 25-03-2022. The study utilises 2014 and 2021 statistics, corresponding to the first and last Code Week editions considered in the study.\\

Adjusted gross disposable income per capita was retrieved from the "Reduced Inequality" section of the Sustainable development indicators \cite{Eurostat:GDP_sdi}. The index is calculated as the adjusted gross disposable income of households and non-profit institutions divided by the Purchasing Power Parities (PPP) of the individual consumption of households and by the total resident population. It reflects households' purchasing power and ability to invest in goods and services or save for the future and consider taxes and social contributions. The dataset records disposable income statistics for EU countries from 2000 to 2019 and was last updated on 17-03-2022. Human Development Index (HDI) was obtained from the Human Development Reports of the United Nations Development Programme \cite{UN:HDI_UNDR}. HDI is a summary indicator of achievements in three critical dimensions of human development:  a decent standard of living, access to knowledge and long and healthy life. HDI is computed as the geometric mean of normalised indices for these three dimensions. The education component is the average of the normalised Expected years of schooling index and the Mean years of schooling index. The health and standard of living index correspond to the normalised Life expectancy and GNI per capita, respectively. The dataset contains HDI measures for all countries worldwide from 1990 to 2019. 
GINI coefficient of equivalised disposable income reported by the EU-SILC survey was obtained from the Social inclusion and social policy indicators section in Tables of EU policy published by Eurostat \cite{Eurostat:GINI}. GINI index is an indicator of statistical dispersion used to represent income and wealth inequality within a nation. The index is defined as the relationship of cumulative shares of the population arranged according to the level of equivalised disposable income to the cumulative share of the equivalised total disposable income received by them. The dataset contains GINI values for European countries from 2009 to 2020. Digital Economy and Society Index (DESI) data published by the European Commission \cite{DA_desi} are available from 2016 to 2021, each based on the latest available data mainly from Eurostat. DESI is a composite indicator with a three-level structure based on data mainly collected from the relevant authorities, aimed at monitoring EU Member States' progress across critical digital areas. The index dimensions are four main interconnected areas that contribute to the country's digital development: Human capital, Integration of digital technology, Connectivity,  and Digital public services. Regional GDP per capita data used in this investigation was produced by the European system of national and regional accounts (ESA 2010) and published by Eurostat \cite{Eurostat:regional}. GDP per capita is expressed in PPS (purchasing power standards) to eliminate differences in price levels between countries. Notably, this indicator is the critical variable for determining the eligibility of NUTS 2 regions in the framework of the European Union's structural policy. The dataset was last updated on 29-03-2022 and reports statistics from 2009 to 2020. It does not include data for NUT 2 regions of the United Kingdom (UK) and Serbia (RS), which won't be considered in the regional part of the analysis.\\

Boundaries of Italian administrative units (cities, provinces, regions and macro-areas) were obtained from the Italian National Institute of Statistics (ISTAT) \cite{IstatData:shapefile}. Updated boundaries are published every year to reflect administrative changes. The most updated shapefile, released on 1st January 2022, was used in this investigation. South Sardinia has experienced the most changes in the last twenty years: it only accounted for the province of Cagliari until 2005, was then split into three provinces (Cagliari, Medio Campidano and Carbonia-Iglesias), and restructured into two (Cagliari and Sud Sardegna) in 2016. GDP per capita estimates are not available for this province due to the frequent changes in administrative units and the incompatibility of different datasets. Population and urbanisation level data was retrieved from the Open Data section of ISTAT and included census data for each Italian city from 2016 to 2021 \cite{IstatData:open}. The degree of urbanisation defined by Eurostat for the Local Administrative Units level 2 (LAU2) classifies them into thinly, intermediate and densely populated areas. These levels are determined using geographical contiguity and a minimum population threshold based on population grid square cells of 1 km$^2$. GDP per capita at the municipal and provincial level was based on income declarations data retrieved from the Department of Finance and Internal Affairs' Open Data section. The study utilises data published in 2015, which refer to the 2014 fiscal year. The income per capita was calculated as the total declared income in a given territory divided by its population \cite{Finance:open}. The average property value was chosen to proxy the socio-economic status at the neighbourhood level. Italian property values were obtained from the Italian Inland Revenue (Agenzia delle entrate) Open Data section \cite{ItalianIRS}. Each municipality is divided into territorial areas (OMI area) reflecting homogenous local real estate market sectors, hence characterised by substantial uniformity in economic and social conditions. The municipal territory is generally divided into zones, which aggregate contiguous and homogenous OMI areas: Central, Semi-central, Suburban 1 (or Peripheric), Suburban 2, Extraurban Zone. For each OMI area of each municipality, the datasets define a min-max price range per unit of surface in euro per square meter, accounting for the state of conservation and the type of property. The study estimates the income level by calculating the mean of the maximum and minimum price for residential buildings.\\


\section{Participation metrics}

This study proposes and calculates participation metrics aimed at providing a comprehensive spatial and temporal statistical description of participation in Code Week. The metrics are defined in Table \ref{tab:def_metrics}, grouped in four main categories: penetration, demographic, retention and spatial metrics. Penetration metrics estimate the number of activities, events creators and hosting institutes in relation to the country's population and the proportion of sub-territories participating in the initiative to quantify the intensity and distribution of Code Week activity. Demographic metrics describe the characteristics of the hosting institute, the organiser and the participants in terms of age, schooling level and gender. Retention metrics assess the continuous participation of creators and locations over the years, estimated as the number of editions per creator (or location) and the proportion of creators (locations) who have joined Code Week for a given number of editions. Spatial metrics estimate the geographical distribution of new events by calculating the number of neighbours a given participant (creator or location) has in a certain area. It also assesses whether new participants occur in a location that has previously hosted events or is close to one. Finally, the participation trend is calculated for all (2014-2021) and pre-pandemic editions (2014-2019) to estimate the impact of the pandemic on the growth rate.

\begin{table}
    \caption{Participation Metrics}
    \label{tab:def_metrics}
    \bgroup
    \def\arraystretch{1.5}%
    \begin{tabular}{p{0.12\textwidth} p{0.3\textwidth} p{0.58\textwidth}}
    \toprule
    Type & Name & Definition\\
    \midrule
    Penetration & Events  Incidence & number of events every 100000 people\\
    & Creators  Incidence & number of event creators every 100000 people\\
    & Locations  Incidence & number of locations hosting events every 100000 people\\
    & Creators Relevance & average number of events per creator\\
    & Locations Relevance & average number of events per location\\
    & Coverage & proportion of provinces hosting at least 1 event\\
    & Total Coverage & coverage calculated in the entire time period\\
    & Average Coverage & average of the coverage calculated in each year\\ \hline
    
    Demographic & Event Size & average of the reported participants number\\
    & Participants Age & weighted average of the reported average participants age for each event, where the weights are the reported number of participants.\\
    & Female Participation & weighted average of the percentage of female participants for each event, where the weights are the reported number of participants. This metric was calculated on all events, and events divided per age group: younger students (age 4-14), older students (14-24), adults (24+) \\
    & Audience Prevalence & percentage of events containing the selected target audience among their target values. This metric was calculated on each target audience and three summative age groups: younger students (pre-school and elementary school children), older students (high school and graduate students), adults (employed and unemployed adults). \\
    & Organiser Prevalence & percentage of events associated to the given type of organiser.\\ \hline
    
    Retention & Creator Recruiting & percentage of new creators in a Code Week editions which have not participated in any previous edition \\
    & Location Recruiting & percentage of new locations in a Code Week editions which have not participated in any previous edition \\
    & Creator Frequency & average number of editions to which a creator has participated \\
    & Location Frequency & average number of editions to which a location has participated \\
    & Creators per \#editions & percentage of creators which have participated in n editions \\
    & Locations per \#editions & percentage of location which have participated in n editions \\
    \hline
    
    Spatial & Neighbours per Creator & average number of creators found in a given area. The area considered is 5 km x 5 km and is identified by a geohash of precision 5\\
    & Neighbours per Location & average number of creators found in a given area\\
    & (new) Creators in previous areas & percentage of (new) creators hosting events in areas which have hosted events in the previous editions\\
    & (new) Creators in near areas & percentage of (new) creators hosting events in areas neighbouring areas which have hosted events in previous editions\\
    
    \bottomrule
    \end{tabular}
    \egroup
\end{table}


\section{Results}

\subsection{Statistical description of Code Week participation in European countries} \label{s_metrics}

\subsubsection{Penetration metrics}

\begin{figure}[h]
    \centering
    \includegraphics[width=0.88\linewidth]{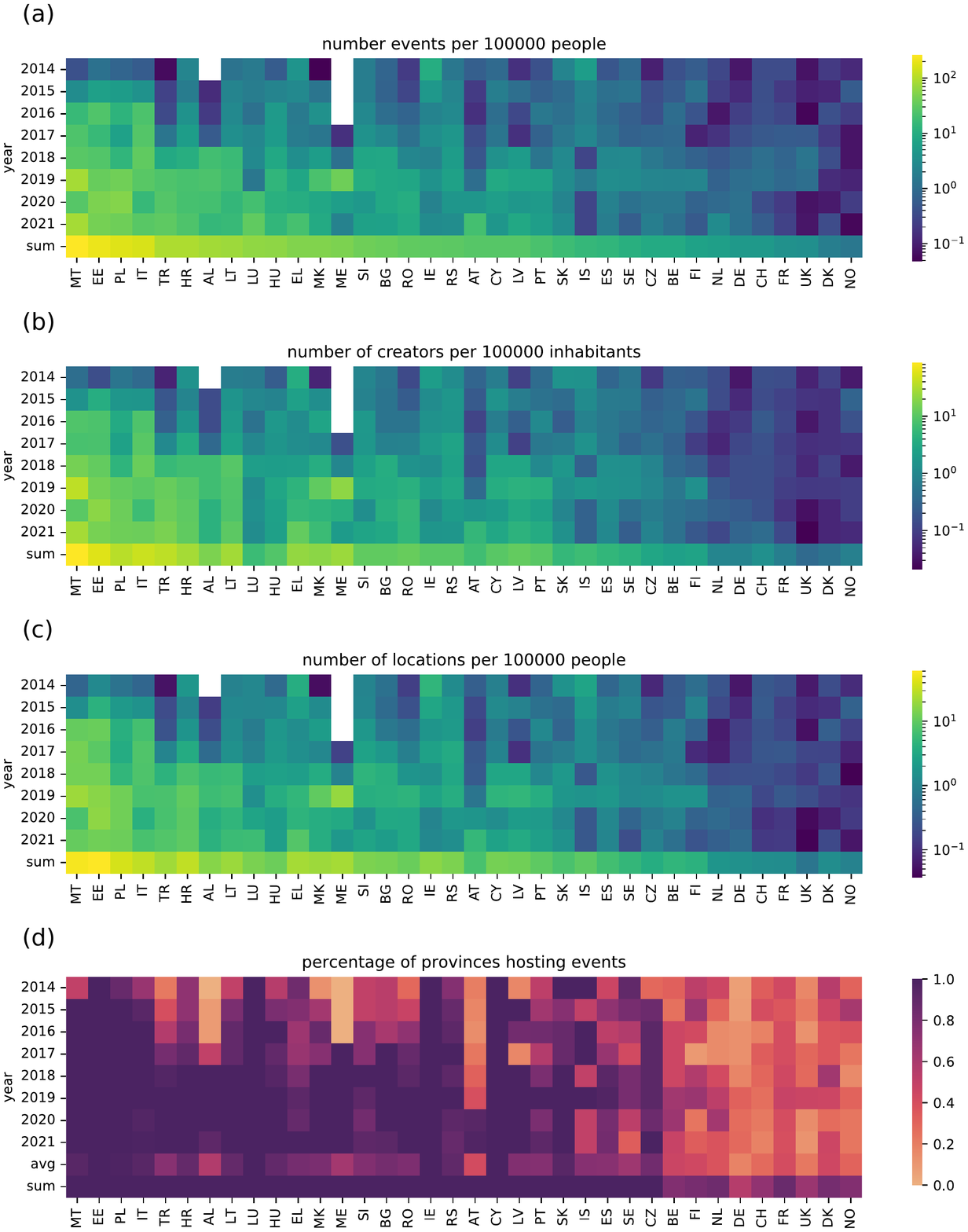}
    \caption{Penetration Metrics}
    \Description{}
    \label{fig:Sm_1}
\end{figure}

Penetrations statistics calculated for EU countries are listed in Table \ref{tab:pen_results}. The selected statistics calculated for each year are shown in Figure \ref{fig:Sm_1}. Entries are sorted by decreasing Event Incidence, and the same order is preserved in all Figures and Tables. In all countries, except Malta (MT), Italy (IT) and Turkey (TR), the number of creators is comparable to or lower than the number of locations. This suggests a  one-to-one or one-to-many correspondence between creators and locations: most creators are the sole participants in their institute, and few organise events in multiple locations. Hence, Code Week's growing participation is not primarily driven by more teachers in participating institutes becoming involved in the initiative but rather by teachers in institutes joining the initiative. On the other hand, in those three countries, the average number of creators per location is above 2, suggesting that teachers also get to know and join the initiative through participating colleagues. A one-to-one creators-location mapping could be explained by the fact that one of the teachers is appointed to create and upload all events for the organisation. However, it is unlikely to be systematic because the average number of events per creator is relatively low (3.8) and similar across countries. Only Luxembourg records more than 10 events per creator, with only 37 creators responsible for all events created in the country and a 3-to-1 ratio between creators and locations. This anomalous trend and the country's low population bring Luxembourg high in the event incidence ranking, but the country's participation is likely overestimated. For these reasons, this country was excluded from the regression analyses in section \ref{s_income}.\\

\begin{figure}[h]
    \centering
    \includegraphics[width=0.88\linewidth]{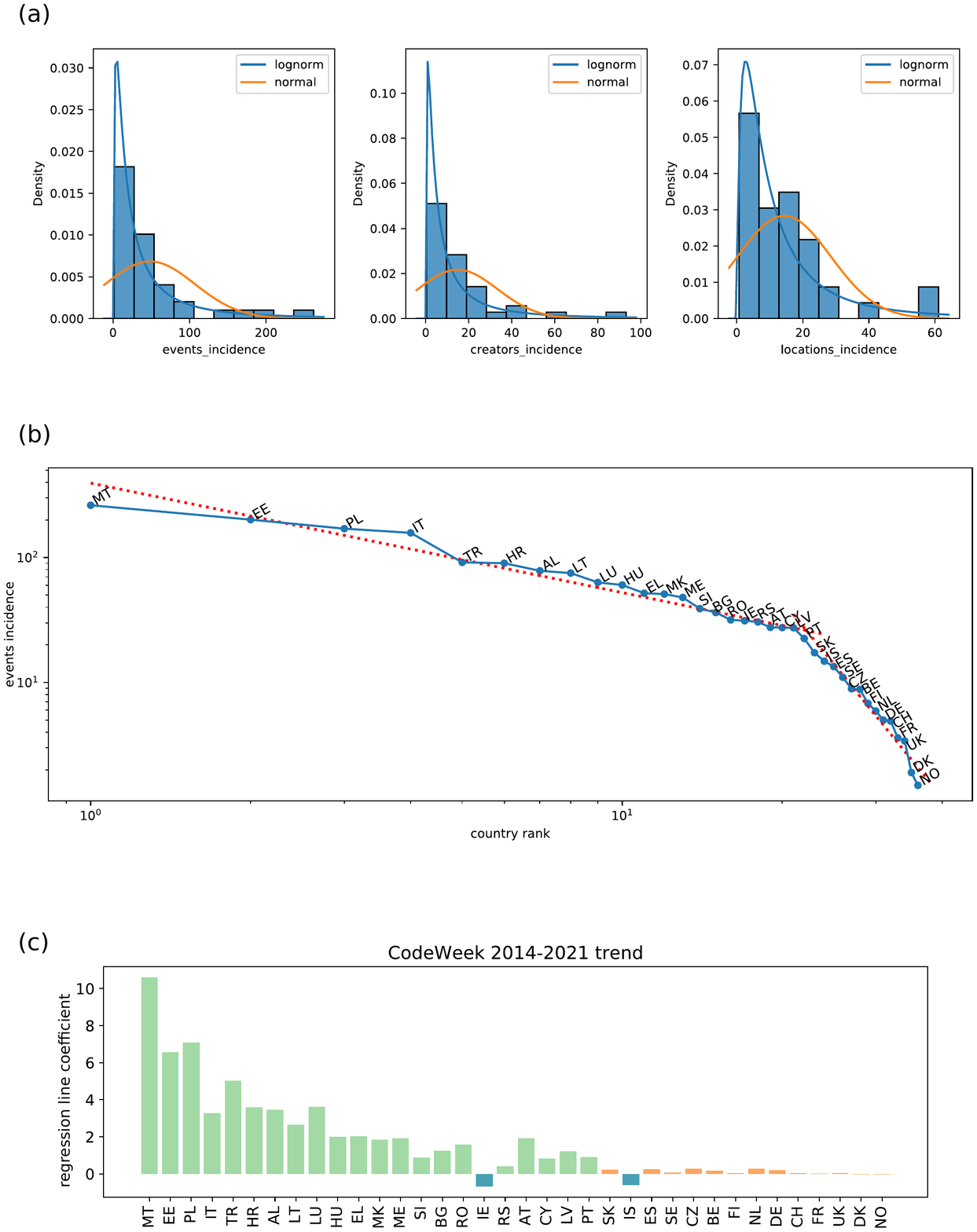}
    \caption{Event Incidence statistics}
    \Description{(a) Distribution of events incidence, creators incidence and location incidence with best-fit normal and log-normal distributions. (b) Event incidence - Rank plot for EU countries. (c) Participation growth trend in 2014-2021}
    \label{fig:fig0}
\end{figure}

Events, creators and locations incidence vary considerably among countries and have a right-skewed distribution, where most countries have medium-low participation while only a handful record very high levels of engagement. The distribution of incidence metrics and the best fit normal and log-normal distributions are shown in Figure \ref{fig:fig0} (a). The latter fits the data well, suggesting that a logarithmic transformation can be applied to normalise the data so that statistical techniques that assume normality and additive effects can be employed. The rank-size plot in Figure \ref{fig:fig0} (b) depicts the distribution of event incidence by the country's rank in decreasing order of participation. This approach is generally used to study data that vary significantly in scale and often distribute according to a power law or other skewed distribution. The rank-size distribution is segmented into two ranges, corresponding to high- and low-participating countries, characterised by distinct behaviour. While the former exhibits a moderate decrease in participation, the latter sees a steep reduction. Subsequent analyses also suggest that European countries have a dual attitude toward Code Week.\\

\begin{figure}[h]
    \centering
    \includegraphics[width=0.88\linewidth]{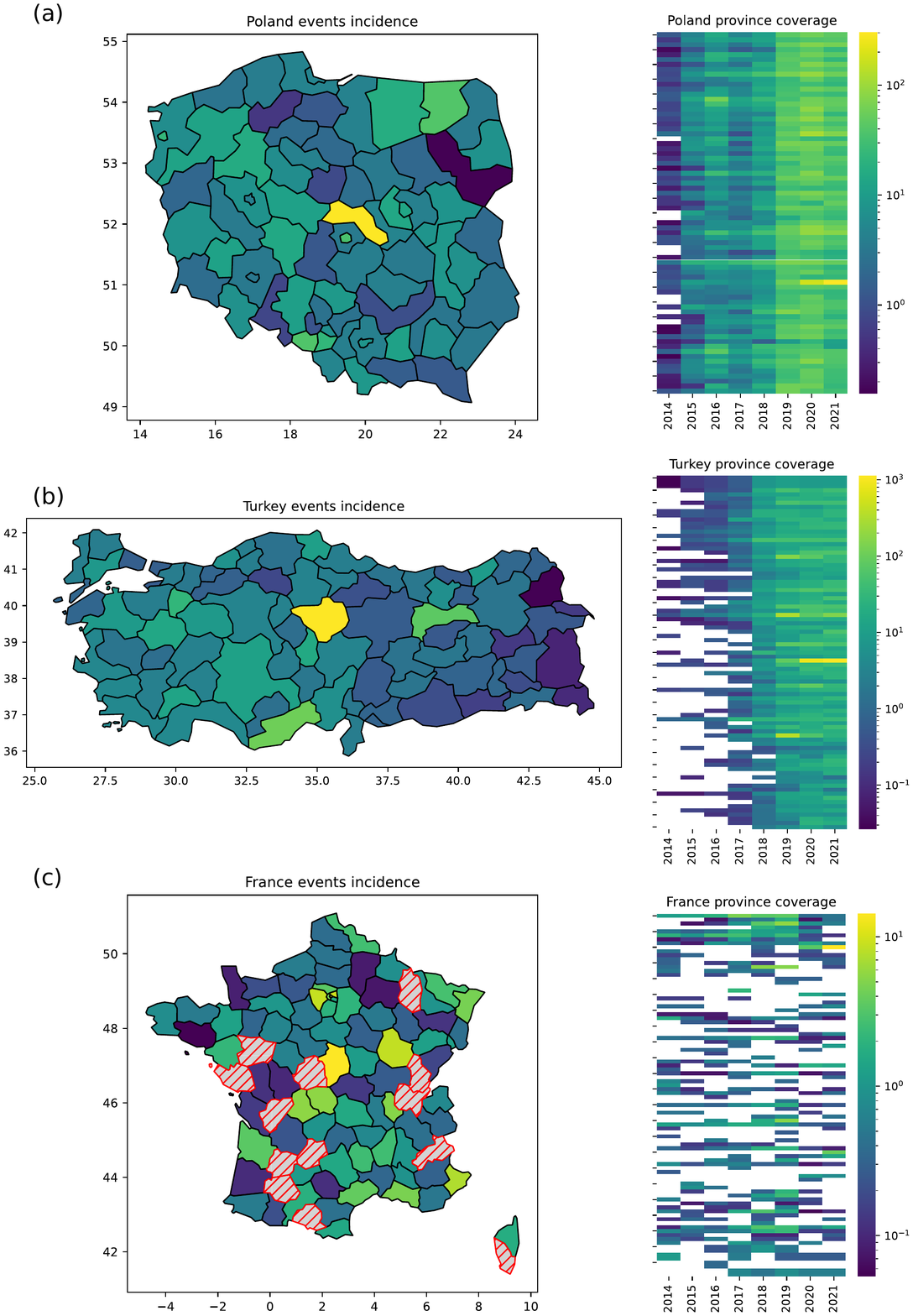}
    \caption{Coverage and total event incidence in three reference countries}
    \Description{}
    \label{fig:Sm_2}
\end{figure}

The total coverage achieved in the last eight editions of Code Week is either complete or close to it for almost all countries. High participating countries generally achieve full coverage immediately or within the first four editions since first joining. Also, when provinces enter the initiative, they continue to do so in the following editions. Low participating countries, however, even when they achieve an overall good coverage (above 70\%), record a low average coverage (below 40\%), meaning that only few provinces are active in each edition and different ones participate every year. In these territories, participation is not only lower in terms of the number of events but also discontinuous. Figure \ref{fig:Sm_2} exemplifies event distribution and coverage of three representative countries with comparable population size and province number, characterised by immediate complete coverage, increasing to complete coverage and low coverage, respectively.\\

\subsubsection{Demographic metrics}
\begin{figure}[h]
    \centering
    \includegraphics[width=0.88\linewidth]{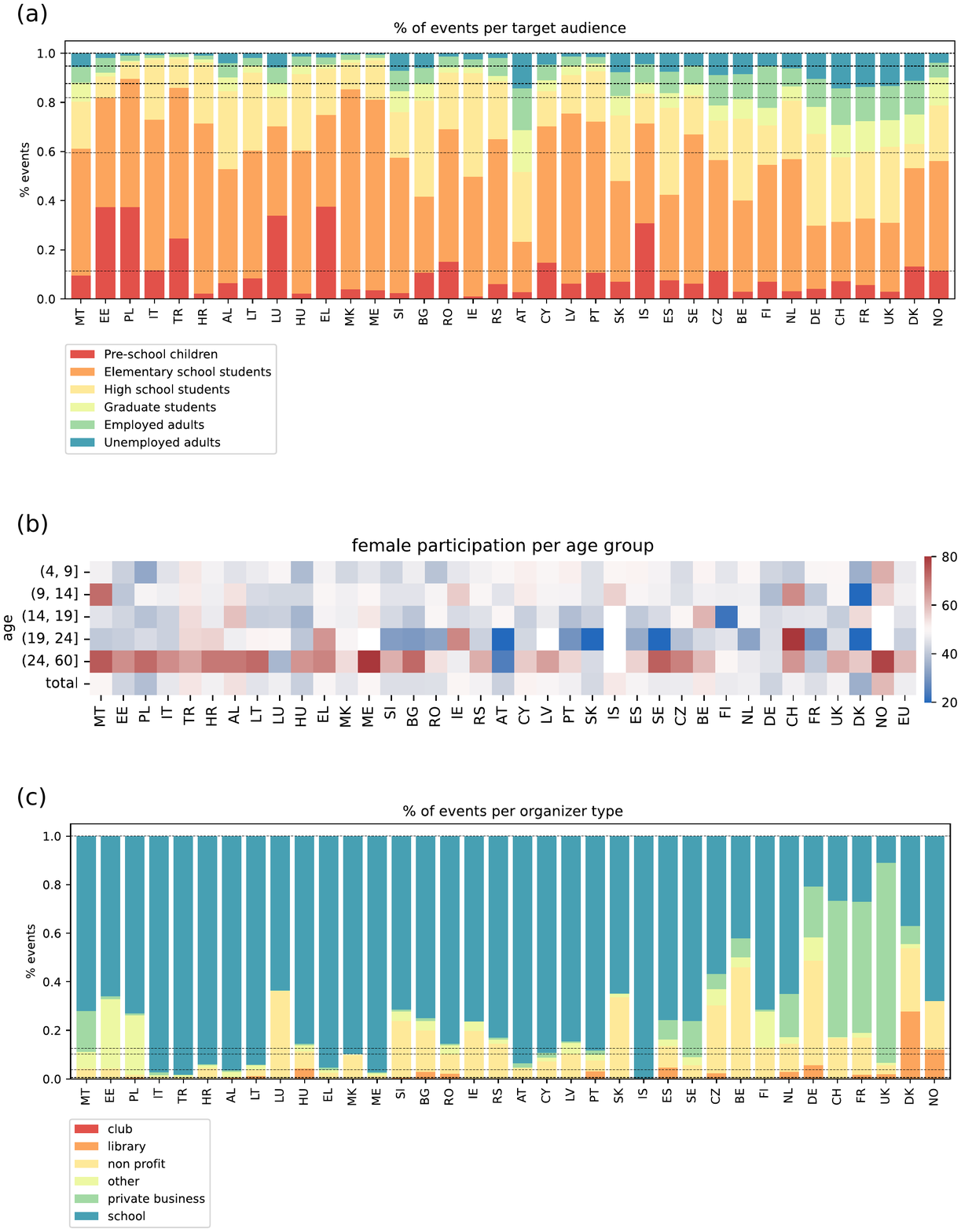}
    \caption{Demographic Metrics}
    \Description{}
    \label{fig:Sm_3}
\end{figure}

Demographic statistics calculated at the country level are in Table \ref{tab:dem_results}, and a more detailed representation of selected statistics is found in Figure \ref{fig:Sm_3}. Demographic data on the participants and the hosting institution are made available through the report, filled by the organiser after the event. Around half of all activities have the report, and the percentage of completed documents appears to be proportional to the degree of participation of the country, with top-participating countries reporting around 70\% of the time, while low-participating countries about 20\%. Completing an expected but not compulsory task is expected to correlate positively with the level of engagement in the initiative. Austria and Luxembourg are exceptions to this rule, recording 4.8\% and 9\% filled reports but medium-high events incidence, suggesting once again that a bulk upload mechanism is in place. \\

The average number of participants per event in different countries varies between 20 and 80, except for the UK, which reports 98. This is likely due to the different organiser composition in the UK with respect to other countries. In top-participating countries, the vast majority of events ( > 80 \%) are organised by teachers and held in a school setting, often during class time. The remaining countries have a more varied composition, with a conspicuous number of events organised by private businesses, then non-profits and libraries (Figure \ref{fig:Sm_3} (c)). The UK records 82\% of events from companies, significantly more than any other country, and only 11\% in schools. Profiles of participation and organiser composition suggest that schools are the driving force of Code Week dissemination. When they are heavily involved, events incidence is high, and the initiative reaches many people. Schools are pervasive and have access to a large group of people and the preferred target audience for the initiative. Companies, however, have limited access to potentially interested people and generally organise specialised events for a dedicated group.\\

The majority of events are directed to primary and lower-secondary students ( $\approx$ 60 \%), followed by high school and graduate students ( 28 \%) and adults ( 12 \%), which reflect Code Week's primary goal to bring computing education to children as early as possible. Low-participating countries (especially DE, CH, FR and UK) have a more significant fraction of events dedicated to unemployed and employed adults and graduate students, mainly due to the larger proportion of events organised by private businesses. Austria (AT) reports the lowest ratio of activities for younger students (23 \%) and the highest for adults (31 \%). Female participation is quite balanced in events dedicated to students, but it grows inversely with the education level. Low female attendance at the university level (age 19-24) observed in most countries reflects the fact that male students disproportionally choose computer science-related degrees. At the same time, events for adults (24+) are predominantly attended by women, likely because these events are workshops, training and professional development activities dedicated to teachers, and most educators, especially at the primary level, are female \cite{femaleteach}.

\subsubsection{Retention metrics}
\begin{figure}[h]
    \centering
    \includegraphics[width=0.88\linewidth]{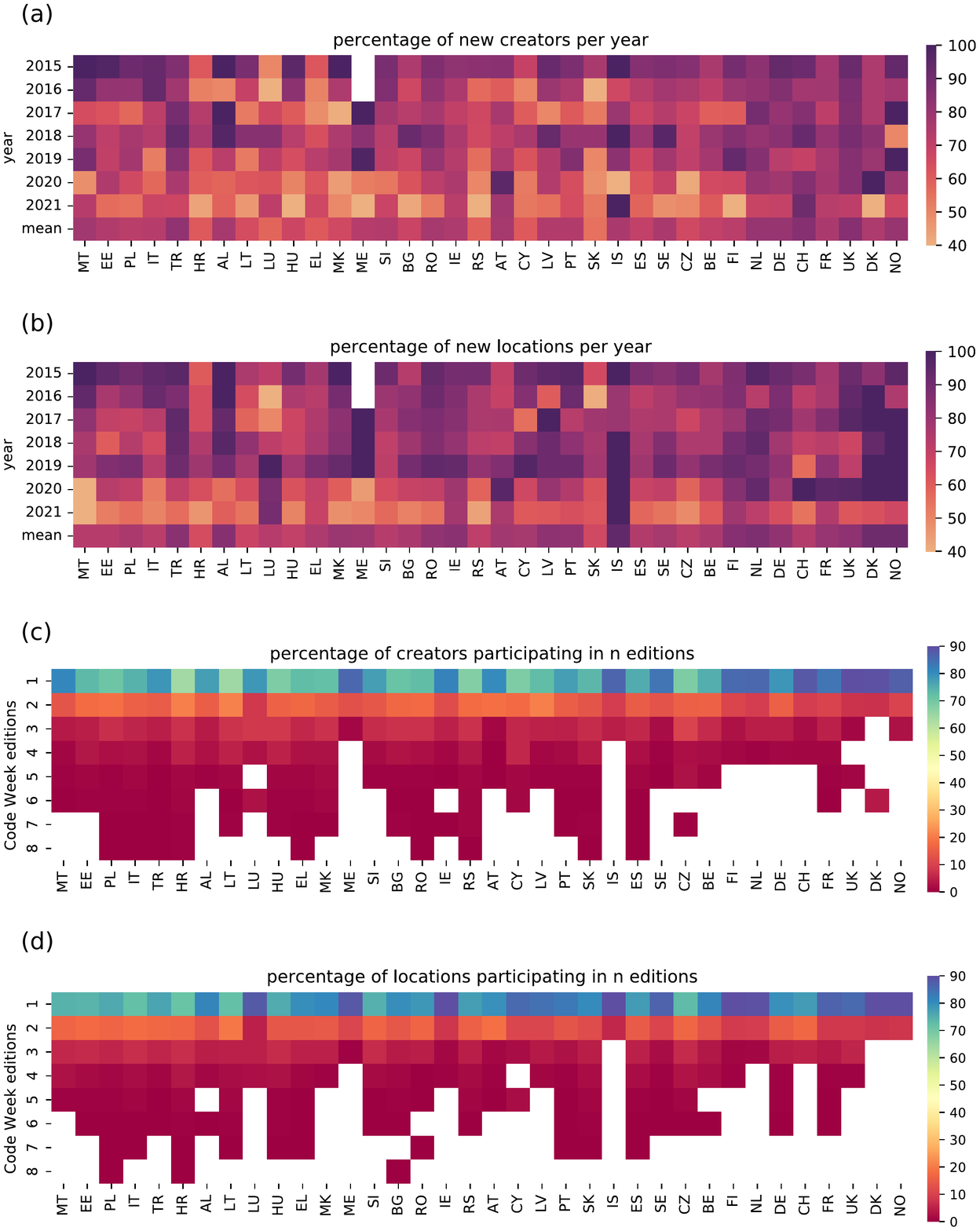}
    \caption{Retention Metrics}
    \Description{}
    \label{fig:Sm_4}
\end{figure}

Retention statistics were calculated for the considered countries and are listed in Table \ref{tab:ret_results}, and yearly metrics are shown in Figure \ref{fig:Sm_4}. Each year about \%72 of the event creators and \%78 of locations have not participated before. These values are consistent across European countries, although they tend to be higher (about 80\%) in low participating countries. The higher turnover of locations is consistent with the one-to-many mapping between creators and locations identified in the previous section, meaning that the same creator organises events in new areas in the following editions. In high participating countries, the percentage of new creators per edition is highest in the first five editions (up to 90\%) but decreases and stabilises at around 50\% in the last three editions. This result suggests a certain degree of saturation, meaning that most people that could have been interested in participating have already been reached, and it is unlikely that subsequent edition will recruit a large proportion of new creators. In other countries, the ratio of new creators is high ( over 90\%) and constant throughout the years, suggesting a fast but stable turnover (Figure \ref{fig:Sm_4} (a)). The same metrics calculated on locations show that the 2019 edition saw an unusually high ratio of new participating areas in most countries. Conversely, the 2021 edition recorded the lowest percentage of new location across Europe (Figure \ref{fig:Sm_4} (b)).\\

The vast majority of creators (77 \%) and location (81 \%) has only taken part in one edition, while only about 8\% of events creators and 5\% of hosting institutes have joined the Code Week campaign for more than two years. Teachers might have registered on the platform multiple times, recording events with different IDs over the years. However, metrics calculated on locations are reasonably similar, suggesting that this behaviour, while possible, is not systematic. In low active countries, around 90\% of creators and locations only join once, coherent with the dispersive and inconsistent participation pattern already observed. In these countries, almost none participate more than 3 or 4 times, indicating low dedication to the cause (Figure \ref{fig:Sm_4} (c) and (d)). \\

\subsubsection{Spatial metrics}

\begin{figure}[h]
    \centering
    \includegraphics[width=0.88\linewidth]{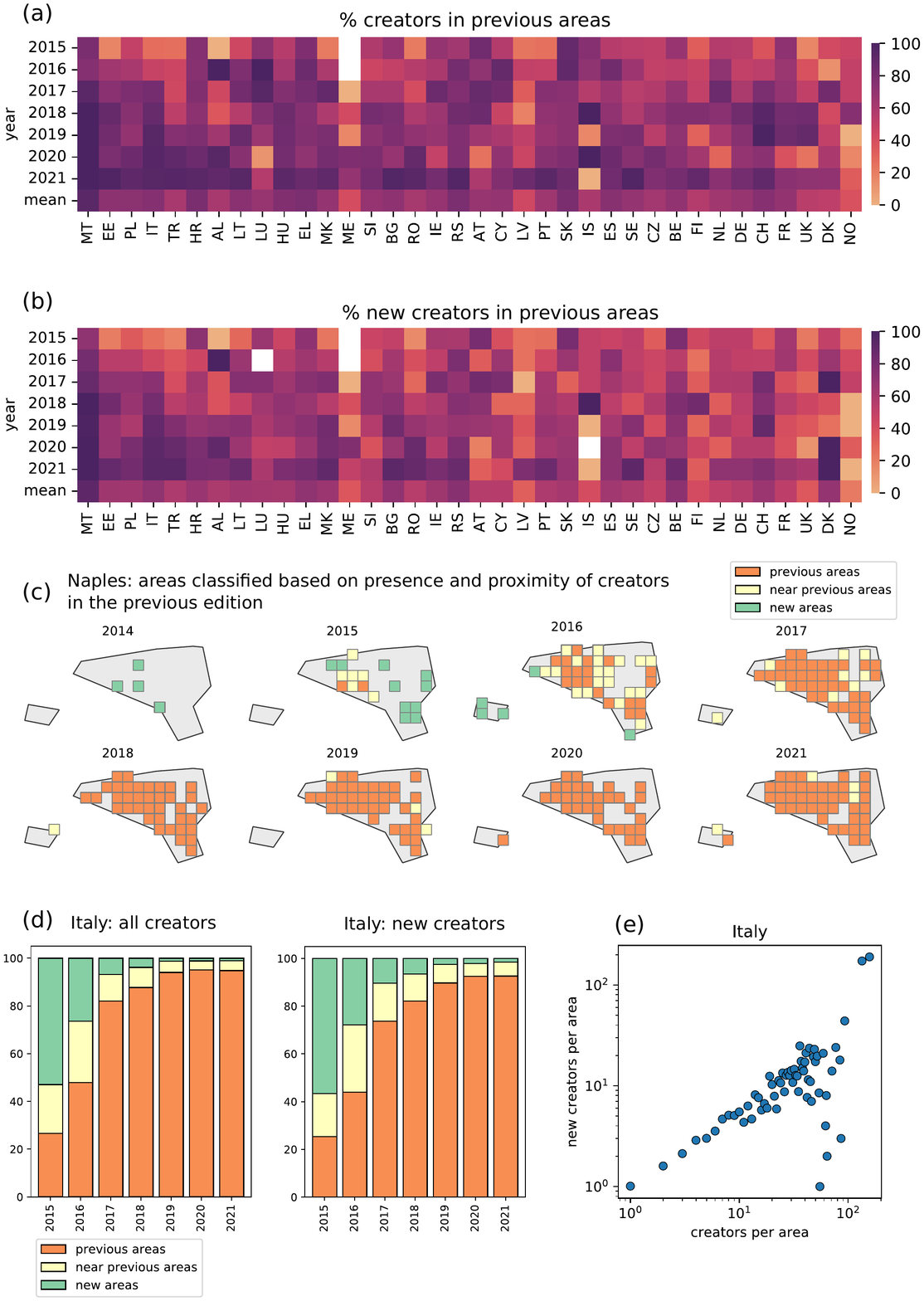}
    \caption{Spatial Metrics}
    \Description{}
    \label{fig:Sm_5}
\end{figure}

Spatial statistics for European countries are listed in Table \ref{tab:spa_results} and relative graphs are in Figure \ref{fig:Sm_5}. Neighbours metrics indicate that areas of 5km x 5km hosting events contain, on average, about four creators and three locations, suggesting that participating creators and institutions are not randomly scattered across the country but tend to form clusters. In Malta (MT), these metrics rise to 25 and 14, respectively, indicating that participation is concentrated in a small area, likely dictated by the demographic and geographic characteristics of the country combined with its elevated participation. \\

The ratio of creators in previous and near locations assesses the percentage of creators found in areas or near areas that have previously hosted events. On average, 67\% of creators organise events in already participating areas, while 13\% in nearby areas. Only 30\% of the creators are responsible for bringing Code Week to new unexplored places. The same metrics calculated solely considering new creators show that more than most new creators occur in previous areas (\% 58) or near them (16\%). This result suggests strong spatial associativity, meaning new creators are more likely to appear when other creators are in the area. The 2021 edition records a higher proportion of creators in previous locations than in earlier years, suggesting a degree of saturation: new creators necessarily occur in areas that have previously participated because most areas have already hosted events, and few are left to reach. Saturation is evident in Figure \ref{fig:Sm_5} (c), which depicts areas active each year in the city of Naples, discriminating among previous, near and new hosting areas. Moreover, Figure \ref{fig:Sm_5} (d) shows a positive correlation between the number of creators in a given area and the number of new creators in that same area, suggesting that the probability of a creator appearing in the area is proportional to the number of creators in that area.

\subsubsection{Trends}

The overall participation trend over the considered period was estimated as the coefficient of the regression line computed on event incidence of a given country over the years. A coefficient between -0.5 and 0.5  indicates the lack of an upward or downward trend but a fluctuation around a specific event incidence value. Trend estimates for EU countries are shown in Figure \ref{fig:fig0} (c). High participating countries exhibit an increasing trend, with a slope coefficient proportional to the event incidence, meaning that the most active countries are also those with the fastest participation growth. On the other hand, less active countries record a stable trend, indicating that they not only participate less but also show no tendency to participate more in the future. Ireland (IE) and Iceland (IS) are the only two countries experiencing a significant decrease in participation in the past decade. The participation growth rate proportional to the level of engagement is responsible for the increasing participation gap between more and less active countries and the skewed distribution of event incidence. Combined with the spatial metrics described in the previous section, these results suggest a mechanism of spatial associativity: people are more likely to participate in the initiative if others have already joined in their proximity, and the more people participate, the more other people will be willing to join too. Simply put, participation fosters more participation within the community, highlighting the role of offline, personal relationships in promoting Code Week. For this reason, Code Week should be assessed in both its civic and social dimensions. A pre-pandemic trend was computed considering only editions from 2014 to 2019, and a difference between the two trends was calculated to estimate a possible "pandemic effect". Pre-pandemic trends are higher for all countries other than Austria (AT) and Norway (NO), suggesting that adopting distance and blended learning and other pandemic measures might have hindered both the motivation and the ability to join Code Week.

\subsubsection{Themes}
\begin{figure}[h]
    \centering
    \includegraphics[width=0.88\linewidth]{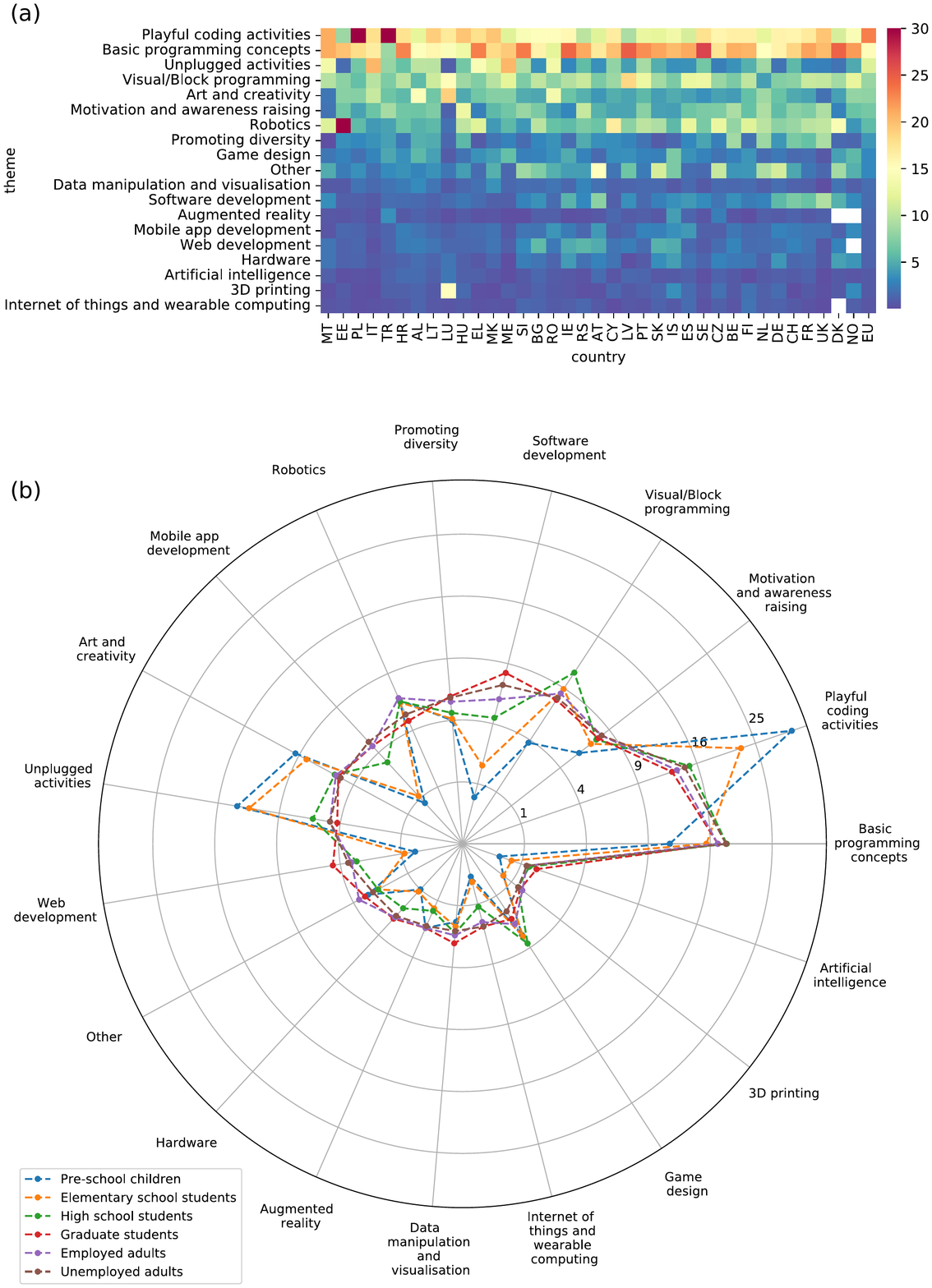}
    \caption{Themes}
    \Description{}
    \label{fig:Sm_6}
\end{figure}

Themes associated with events provide further information on the tools utilised (e.g. Visual/Block Programming, Unplugged Activities), the skills acquired (e.g. Basic Programming Concepts, Software Development), and the aims achieved, namely Motivation and Awareness, and Promoting Diversity, through the Code Week activity. These last two themes are the 6th and 8th most chosen themes and are evenly popular across countries and target audiences, confirming Code Week value as an awareness-raising campaign other than a coding education provider. The distribution of themes tends to be similar across countries, as shown in Figure \ref{fig:Sm_6}, with more popular general themes (Playful Coding, Basic Programming Skills, Unplugged Activities) and less frequent specialised themes (Artificial Intelligence, 3D Printing, Internet of Things). As previously observed, in less active countries, more events are organised by private businesses for an adult target audience, which is reflected in fewer Unplugged Activities events in favour of more Software Development ones. Events dedicated to a younger audience (i.e. primary and lower-secondary education classes) favour Playful Coding, Unplugged Activities, and Art and Creativity, reflecting Code Week's goal 
to bring coding education to children in an accessible and enjoyable way. Events intended for adults focus on developing both general and specialised skills, such as Robotics and Mobile App Development.

\subsection{Code Week participation and Income}\label{s_income}

\subsubsection{EU countries}

This section investigates the relationship between Code Week participation and the socio-economic status in European territories. The event incidence is chosen to proxy the participation level of a given area. It has the highest variability among all metrics; thus, it can better highlight differences in participation among countries. A logarithmic transformation was applied to this indicator before regression due to its skewed nature, as shown in Figure \ref{fig:fig0} (a). The normality test confirmed that the transformed index could not be distinguished from the corresponding best fit normal distribution (p-value > 0.05). \\

\begin{figure}[h]
    \centering
    \includegraphics[width=0.88\linewidth]{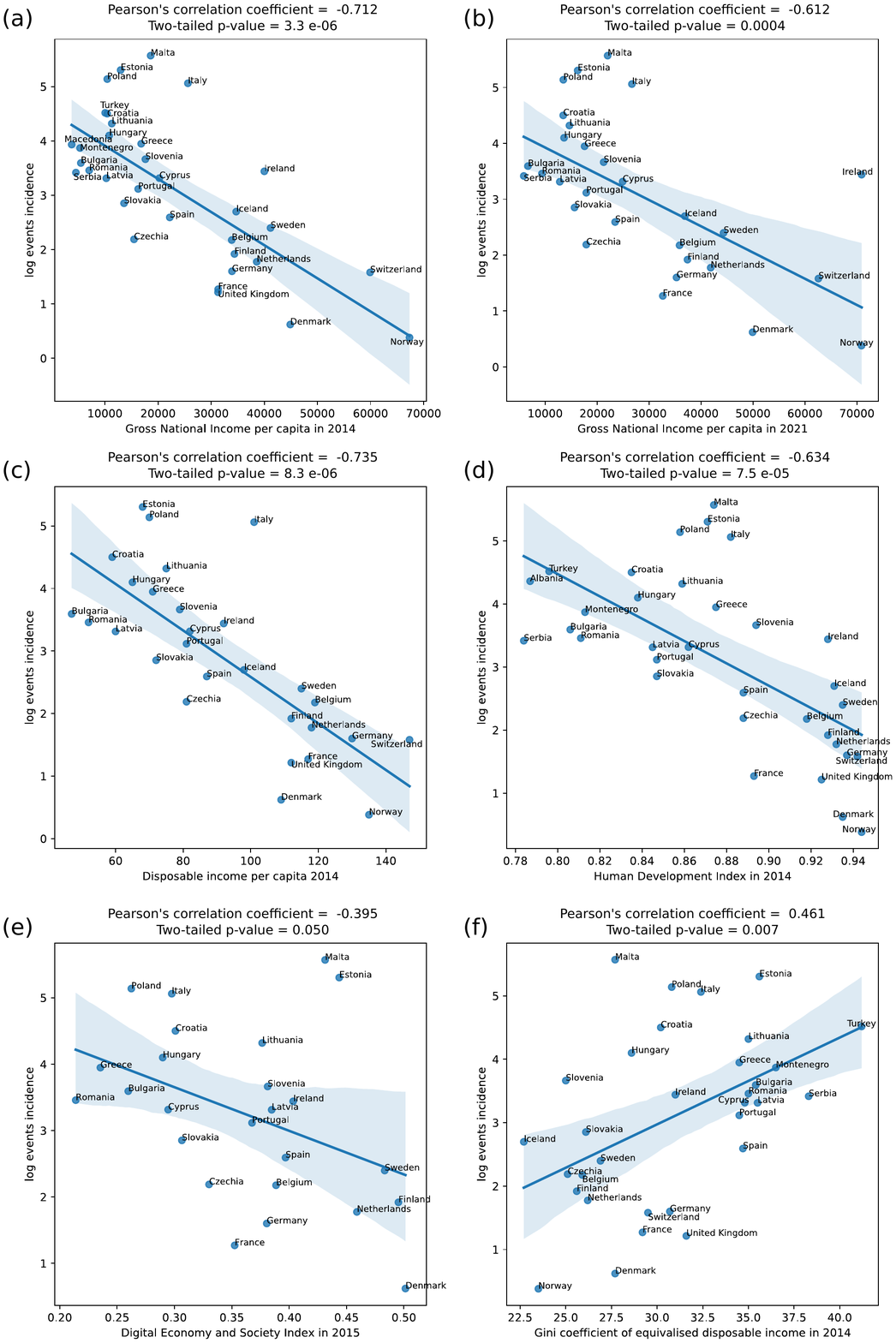}
    \caption{Correlations of events incidence with socio-economic metrics}
    \Description{}
    \label{fig:fig1}
\end{figure}

Regression on Gross National Income per capita for 2014 and 2021, corresponding to the first and last Code Week editions considered in this investigation, shows an inverse linear correlation with participation. Correlation is strong and significant in 2014, with a Pearson correlation coefficient = -0.712 and a p-value < 0.001 (Figure \ref{fig:fig1} (a)). Dispersion is higher, and correlation is lower in 2021 (Figure \ref{fig:fig1} (b)), although still considerable and significant, mostly due to the change in Gross National Income per capita occurred in Ireland, which increased by 78\% from 2014 (40010) to 2021 (70920), whereas all other European countries experienced an increase between 4\% and 30\%. The previous section showed how the participation trend in the first editions strongly affects future participation, i.e. countries heavily or mildly involved in the first four years continue to do so in the subsequent four years. Ireland constitutes the only exception, having a diffused and intense participation at the beginning and a significant decline toward the end. Consequently, participation should be analysed in relation to the country's socio-economic status at the beginning of the considered period (2014-2015). \\

Other development indicators were included to narrow down the socio-economical factors affecting events incidence and begin to unveil the underlying mechanisms influencing participation. Disposable income is more suitable than GDP per capita to estimate the resources available to the citizen, as it directly relates to the households' purchasing power and ability to invest and save for the future. Disposable income in 2014 shows a stronger correlation than the GDP per capita in the same year (Figure \ref{fig:fig1} (c)), with a Pearson correlation coefficient of -0.735 and p-value < 0.001, indicating that participation is higher when the average citizen has fewer resources available to them. Italy is the only country that significantly deviates from this trend, exhibiting higher participation than expected. The heterogeneous development level that characterises this country, described in detail in the next section, might explain this anomaly. \\

A country's GDP strongly determines the quality of its education system. It can be argued that the GDP affects participation through the infrastructure and services that the government provides to its citizens. For instance, citizens of regions with lower GDP might participate more in Code Week because they do not benefit from high-quality education and are looking for further opportunities elsewhere. The Human Development Index (HDI) is a more suitable measure of a country's ability to offer its citizens a good standard of living. This indicator combines an economic dimension with a health and education dimension to emphasise that development should consider people and their capabilities, other than economic growth. It allows questioning government national policies and priorities when two countries have comparable GDP per capita but very different HDI. HDI also displays a negative correlation with event incidence with a Pearson correlation coefficient of -0.634 and p-value < 0.001 (Figure \ref{fig:fig1} (d)). The four most participating countries, i.e. Malta, Estonia, Poland and Italy, deviate from this trend, having the highest event incidence and relatively high HDI. However, when controlling for income, the effect of HDI on participation is not significant (p-value = 0.976), indicating that the development degree of a country's population does not mediate the relationship between income and participation. On the contrary, GDP per capita is likely to confound both variables.\\

A country's GDP strongly correlates with the ability of its citizen to access and use technology (e.g. computers, smartphones, internet broadband). Consequently, the effect of GDP on participation in digital literacy campaigns could be mediated by the country's digital infrastructure and the population's digital capacity. Under this hypothesis, citizens of less developed regions would be more prone to participate because they have had less opportunity to use and learn about technology. DESI is a composite indicator that captures the overall digital performance and competitiveness. Its correlation with events incidence is significant (p-value = 0.050), with a Pearson correlation coefficient of -0.395 (Figure \ref{fig:fig1} (e)). Notably, Malta and Estonia have a much higher degree of digitisation with respect to their GDP, mostly because of targeted government policies, but are also heavily involved in Code Week. Similarly to HDI, the digital economy and society index does not significantly affect participation when controlling for GDP per capita (p-value = 0.681), disproving the digital infrastructure's mediating role on participation.\\

Most studies on the impact of socio-economic status on civic engagement identify social and economic heterogeneity as a determining factor. Income inequality is known to negatively correlate with the average national income, meaning that more developed countries also tend to distribute their wealth more equally \cite{kim2016study}. The GINI coefficient of equivalised disposable income is the most common indicator to quantify wealth inequality. It correlates positively with events incidence, as one would expect, with a moderate correlation coefficient of 0.461 and p-value = 0.007 (Figure \ref{fig:fig1} (f)). However, the relationship is not significant when accounting for GDP per capita (p-value = 0.690).\\

Many socio-economic statistics are available for European countries through Eurostat, and many more correlations can be easily explored. However, spurious correlations are likely to appear because of the strong correlation of GDP with participation and most socio-economic metrics, where GDP acts as a confounding variable. \\

\subsubsection{EU regions}

\begin{figure}[h]
    \centering
    \includegraphics[width=0.88\linewidth]{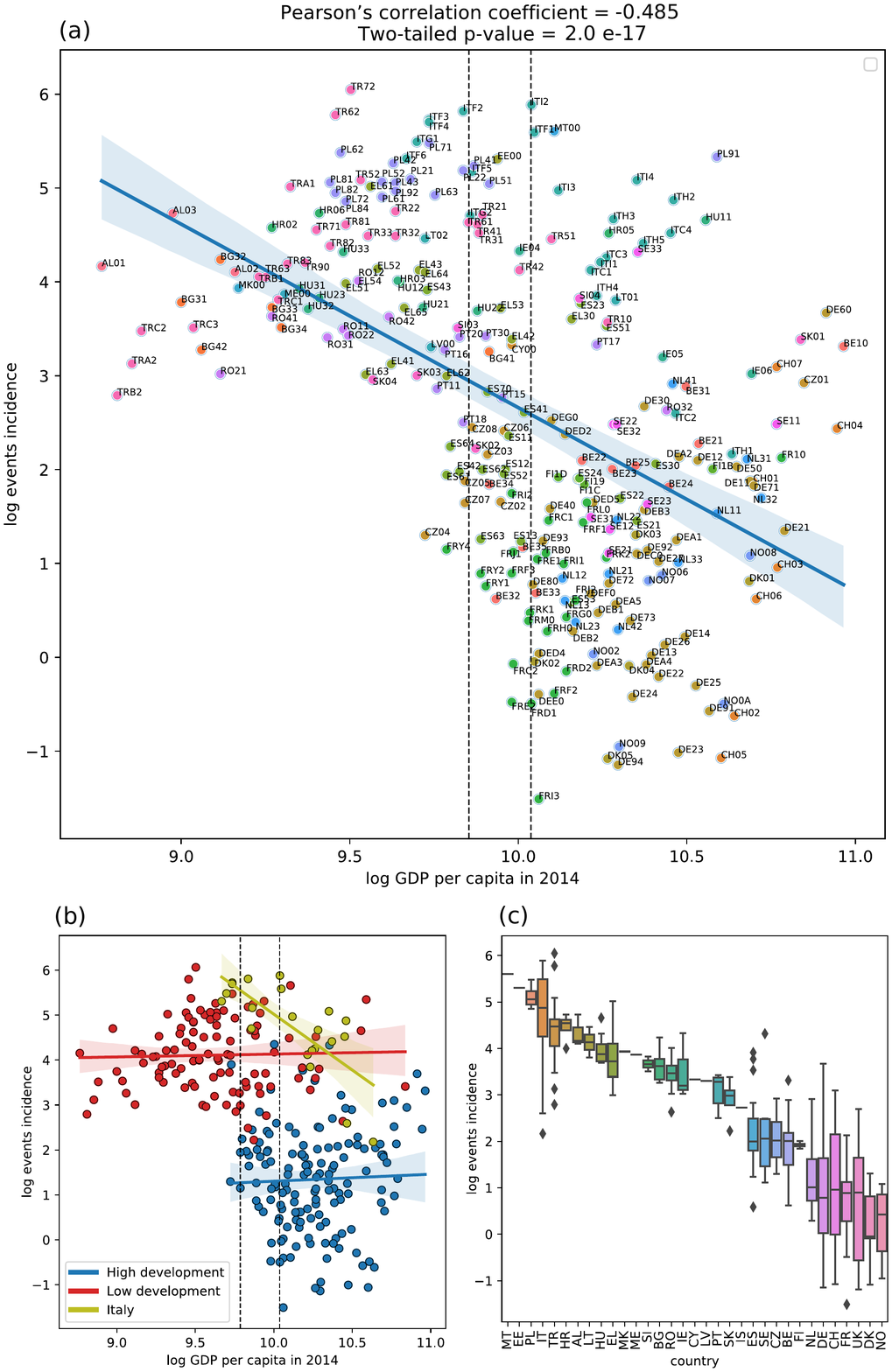}
    \caption{Relationship between Regional GDP per capita and events incidence with regression line on (a) all regions (b) regions divided by development level of the country. (c) Mean and standard deviation of regional GDP in EU countries.}
    \Description{}
    \label{fig:fig2}
\end{figure}

The relationship between participation and GDP was then investigated at the regional level to evaluate the participation distribution within countries and verify whether the correlation identified in the previous section persists at a smaller scale. GDP per capita was plotted against events incidence, both in the logarithmic domain due to their highly skewed nature, as shown in Figure \ref{fig:fig0} (a). Linear regression confirmed the negative linear relationship found at the country level with Pearson's correlation coefficient of 0.485 and p-value < 0.001 (Figure \ref{fig:fig2} (a)). The missing GDP per capita data for the UK and Serbia does not allow for direct comparison with the country-level data but confirms the trend.\\

Less developed countries are characterised by high participation levels in all regions. Participation remains close to the country average, even when the GDP is vastly different. For instance, in Poland, all territories except one have a GDP per capita in the 13000 - 20000 (PPS) range and participation between 120 and 240 events per 100000 people. The Warsaw region, which hosts the country's capital, claims more than double the GDP per capita (39700) but retains a high events incidence of 206. A similar trend is observed in developed countries, where the GDP per capita is consistently high and participation is low across all regions. Less developed areas in otherwise wealthy countries are too characterised by inconsistent and sporadic participation.\\

The graph includes two fundamental GDP thresholds, 75\% and 90\% of the average European Union GDP, which determine the eligibility of regions in the framework of the European Union's structural policy. Regions with a GDP per capita of less than 75\% of the EU average are designated as less developed and assigned the most significant regional policy funding through the Regional competitiveness and employment objective. Regions whose GDP per capita falls between 75 and 90\% are instead considered transition regions and have access to less funding than the less developed regions but more than developed regions. Notably, almost all regions classified as less developed, i.e. falling behind the 75\% mark, record very high event incidence.\\

EU countries are relatively homogeneous with respect to this classification. Most EU's "old" Member States (AT, BE, DK, FI, FR, GR, IE, LU, NL, ES, and SE) are characterised by all developed regions or a minority of transition regions. In contrast, the new EU Member States (BG, HR, CZ, EE, HU, LT, LU, PL, RO, SK, SI), together with the old members Portugal (PT) and Greece (EL), consist of a majority of transition and less developed regions, and only a minority of developed regions, generally corresponding to the capital area. Italy (IT) is unique in this regard, as it appears to be split into three parts. Its 2014-2020 EU regional policy classification has five less developed regions, which account for $\approx$ 30\% of the population, three transition and 12 developed regions. Thus, the Italian situation requires individual consideration. \\

Figure \ref{fig:fig2} (b) shows the same relationship between GDP and events, but with regions classified based on the development level of the afferent country. In high and low development countries, participation almost exclusively falls within the 0-20 and 20-400 event incidence, respectively. Within a given country, participation is comparable across regions, even in the case of variations in GDP, suggesting that the average development level of the country rather than the GDP of the particular area affects participation. Italy constitutes an exception, as it comprises two territories with vastly different socio-economic profiles. Participation is high overall in the entire country, but more developed areas have significantly lower events incidence than transition and less developed ones. \\

Figure \ref{fig:fig2} (c) confirms that variability in participation within a country is relatively low in high participating countries, except for Italy, due to its dual nature. OLS regression model with GDP per capita as the independent variable had very little explanatory power (R-squared = 0.28). A significantly better model (R-squared = 0.64) was obtained when adopting a discrete development level variable (1: High, 2: Medium, 3: Low) as the independent variable. Adding the GDP attribute did not improve the model's accuracy, for which only the development variable had a significant p-value. Figure \ref{fig:fig3} (a) shows GDP per capita and events incidence to highlight the complementarity of the two variables. \\

\subsubsection{Italy}

\begin{figure}[h]
    \centering
    \includegraphics[width=0.88\linewidth]{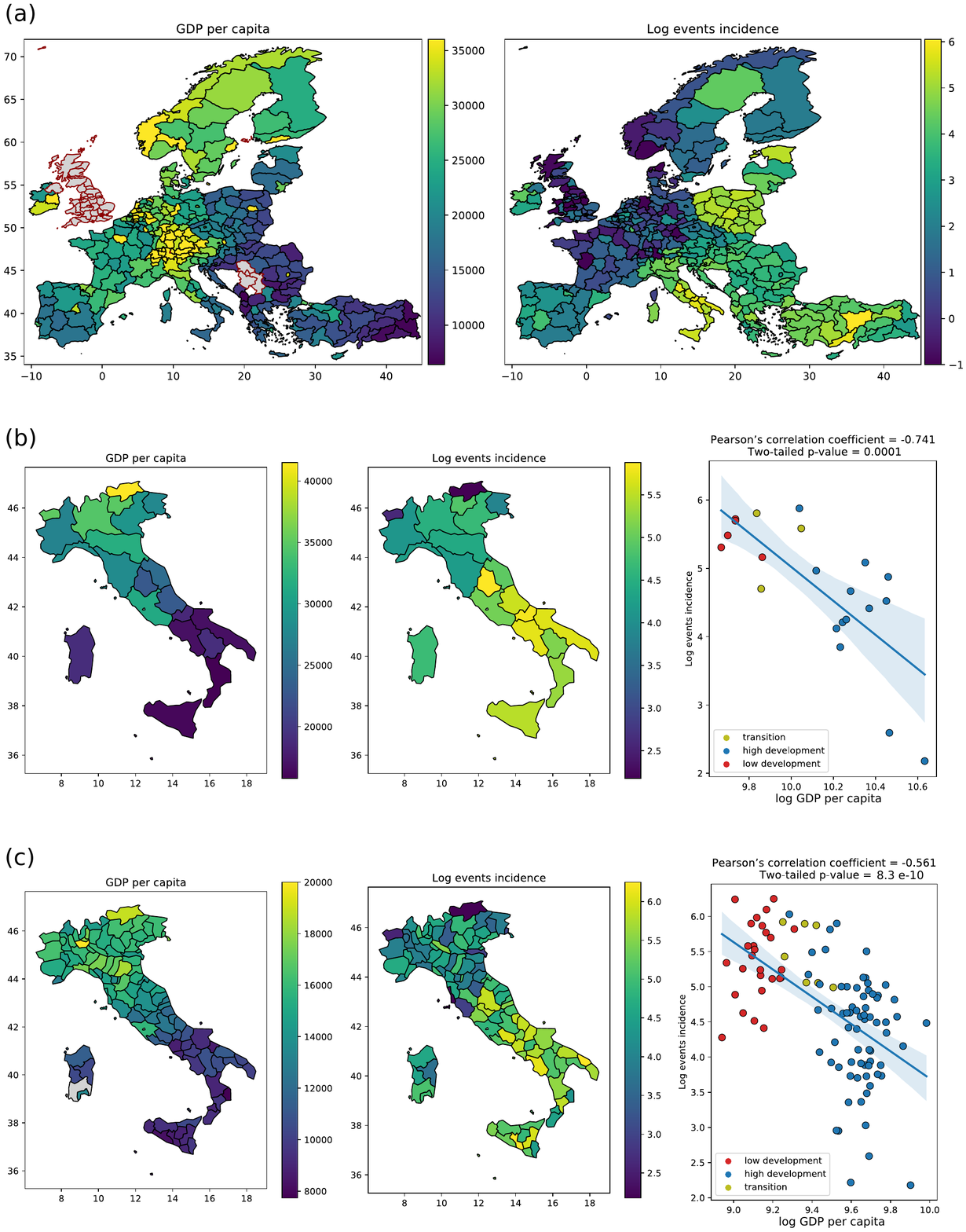}
    \caption{Relationship between GDP per capita and events incidence in (a) European regions, (b) Italian regions, and (c) Italian provinces}
    \Description{}
    \label{fig:fig3}
\end{figure}

Given its heterogeneous nature and atypical participation pattern, the Italian situation deserves further consideration. Regression between regional events incidence and GDP per capita highlights a negative linear correlation, similar to that observed at the European level, with a strong Pearson's coefficient (0.74) and a significant p-value (Figure \ref{fig:fig3} (b)). All transition and low-development regions are heavily involved in Code Week, with over 50 events every 100000 people, while participation is more varied in high development regions. This pattern is similar to that observed at the country level, where less developed areas consistently recorded high events incidence. In contrast, higher developed areas had lower events incidence but higher variance due to a minority of wealthy but participating regions. \\

The same analysis was repeated at the province level to study the participation distribution within a region and yielded similar results (Figure \ref{fig:fig3} (c)). The negative linear correlation has a moderate Pearson's coefficient (0.56) and a significant p-value. Once again, participation tends to be homogeneous in provinces within a region. All provinces in transition and less developed areas have consistently high events incidence, while more developed areas vary into a broader range of participation. Events distribution was then analysed with respect to the urbanisation level of the hosting municipality. Thinly, intermediate and densely populated areas hosted 15\%,  50\%, and 35\% of the events, respectively, suggesting that Code Week was not confined to large cities but was significantly present at each urbanisation level and was most popular in medium-sized settlements. \\

This section highlights a robust negative correlation between event incidence and GDP at the country, region and province levels, suggesting that lower-income individuals are more prone to participate. However, it could be possible, albeit unlikely, that participation is concentrated in the wealthier parts of the cities, driven by high-income individuals living in an overall low-income setting. Participation was further investigated at the neighbourhood level to exclude this possibility and draw conclusions on participation trends at the individual level. The property value of the hosting institution was used to proxy the income level of organisers and participants \\

\begin{figure}[h]
    \centering
    \includegraphics[width=0.88\linewidth]{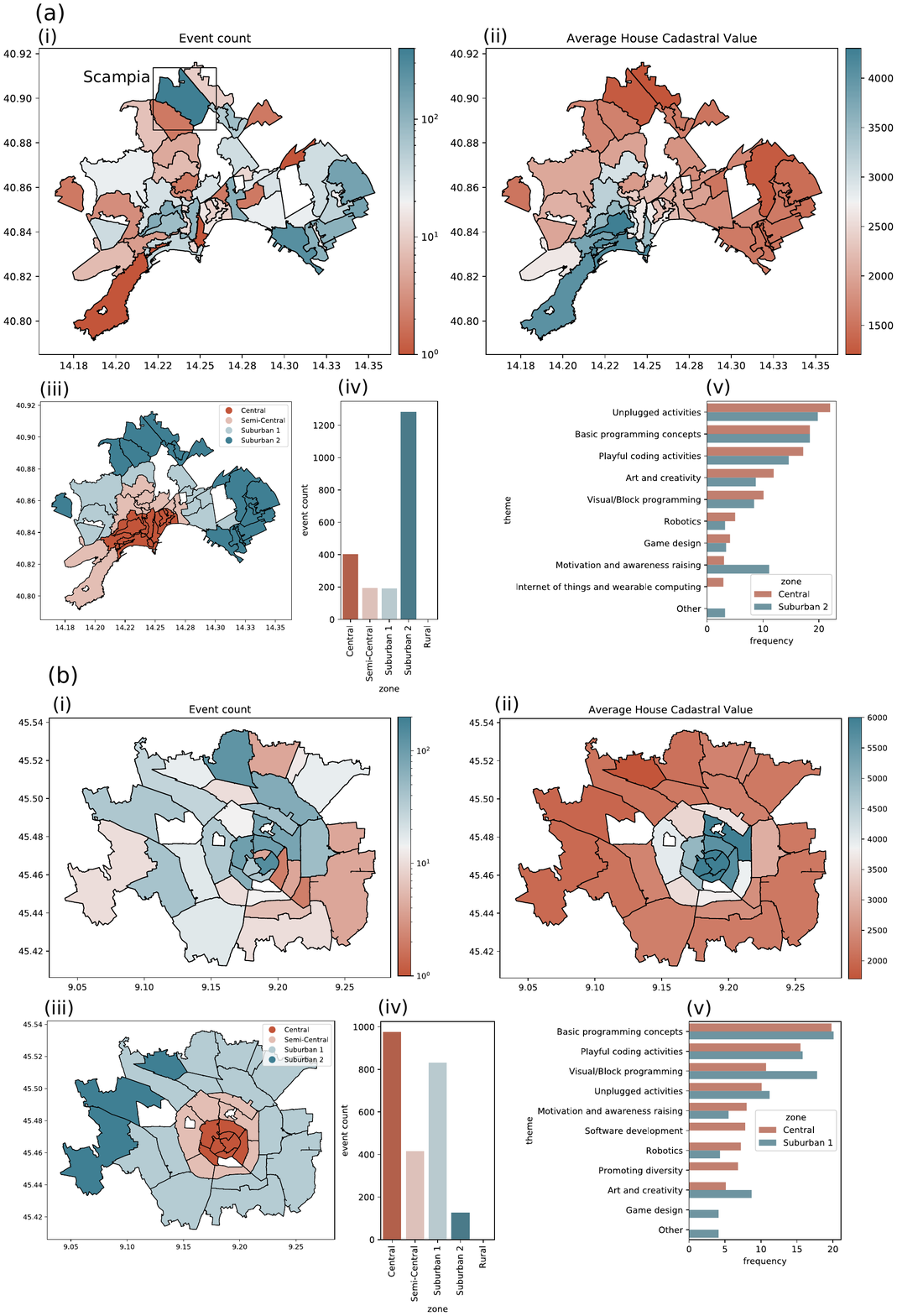}
    \caption{Income and participation statistics for two reference Italian cities, (a) Naples and (b) Milan: (i) event count, (ii) property value, (iii) municipality zones, (iv) event distribution per zones, (v) themes frequency}
    \Description{}
    \label{fig:fig4}
\end{figure}

Events distribution was further analysed in two reference cities, Naples and Milan, respectively, representing the country's less developed south and the more developed north. The events distribution in Naples in Figure \ref{fig:fig4} (a) (ii) shows that almost every neighbourhood has hosted events and that the most active areas correspond to those with the lowest property value. Notably, the Scampia neighbourhood, highlighted in Figure \ref{fig:fig4} (a), records the lowest average property value at X and the highest number of events at Y. According to the OMI zone classification, shown in Figure \ref{fig:fig4} (a) (iii), over 62\% of all events were mapped in suburban neighbourhoods, while only 19\% occurred in the central areas, as shown in Figure \ref{fig:fig4} (a) (iv). The themes of events in these two zones, characterised by opposite socio-economic profiles, are relatively similar, except for a more significant proportion of events focused on "Motivation and awareness-raising" in the suburbs (Figure \ref{fig:fig4} (a)(v)). In Milan, all considered neighbourhoods have participated at least once, but activities seem to be concentrated in the city's centre, having the highest property value(Figure \ref{fig:fig4} (b) (i-ii)). The central zone hosts 43\% of the events, followed by the peripheral (35\%), semi-central (17\%), and suburban(5\%) (Figure \ref{fig:fig4} (b) (iv)). Themes are fairly similar in the two most participating zones, central and peripheral, with a preference for "Promoting diversity" and "Software development" in the centre (Figure \ref{fig:fig4} (b) (v)). \\

\section{Discussion}

The participation metrics calculated in this study aimed at evaluating the reach of the Code Week campaign across Europe in the past eight editions and investigating the underlying participation mechanisms, discriminating between traits common to all countries, hence endogenous to the Code Week program, and features varying across countries, likely to be influenced by factors internal to the respective country. The metrics aim to describe both incidence and amount of participation, i.e. whether an individual chooses to participate (in this case to organise and activity) and how often, while most empirical studies in the literature only address one aspect \cite{schroder2021socio}. Coverage statistics indicate that most European countries have hosted events and that in over two-thirds of all considered countries (25 out of 36), all provinces have participated at least once, and most take part in every edition. Spatial metrics calculated in the last editions with respect to the earlier ones indicate a degree of saturation, particularly in highly participating countries, meaning most areas suitable to host events have already joined the campaign. Code Week appears to have reached and become well known in most European territories, confirming its success as an awareness-raising campaign. \\

Events incidence, together with creator and location incidence, varies considerably among countries, ranging between 1 and 250 events every 100000 people. The gap between high and low participating countries is large and expected to widen since the former countries are also characterised by a fast participation growth rate, while the latter fluctuate about a fixed value. High participating countries have a continuous Code Week presence on the territory. The number of active members within an area increases even though the turnover is high, and members appear to form clusters, suggesting the presence of offline, personal relationships promoting Code Week dissemination. On the other hand, in less active countries, participation lacks both spatial and temporal continuity, indicating the lack of an underpinning social structure. Code Week recognised the importance of having active members on the territory and established a net of ambassadors and leading teachers across Europe. However, results suggest that relationships at the municipality or neighbourhood level might be just as relevant in fostering participation. Efforts should be directed toward building and maintaining a dense network of close-range interaction, for instance, promoting group activities to be completed by different classes of the same or neighbouring institutes.\\

All countries report relatively low creator and location relevance and low retention levels, suggesting that creators generally join Code Week with few events and only a few editions. Nevertheless, the number of Code Week events is overall increasing in almost all countries, indicating that new creators joining each year are the driving force to participation, rather than previous creators remaining active in the program. These results suggest that Code Week is more effective in recruiting new teachers to the program than retaining them. Efforts should be made to maintain contact with previous participants and inquire about possible dissatisfaction. Reasons leading teachers to discontinue their participation should be further investigated, and retention measures purposefully implemented. Demographic metrics and themes tend to be homogeneous across countries, except for the type of event organiser and the target audience. In low participating countries, a more significant portion of events is organised by private businesses, generally for an adults audience, and generally on a specialised IT topic, while in other countries, activities are almost exclusively managed by schools for young students. Demographic data found in reports is valuable in assessing the campaign's effectiveness but is only available when volunteered by the teachers and is rare in low participating countries. Incentives could be offered to increase reporting and obtain more comprehensive and representative demographic data.\\

The article also investigates the socio-economic determinants of participation in the digital literacy campaign EU Code Week. The analysis identified linear correlations at the country level with the socio-economic parameters relative to 2014 and found GDP per capita and disposable income to be most relevant. Investigating population development indices (HDI and DESI) while controlling for GDP per capita demonstrated how income does not primarily affect participation through the educational, technological, health, and social infrastructure it provides to the population but rather directly through the availability of resources. The strongest correlation is, in fact, observed with disposable income, an indicator designed to reflect the average households' purchasing power and ability to invest in goods and services and adequate to estimate the resources available to the citizen. Similarly, studying the relationship with income inequality (GINI) while controlling for GDP highlighted that the overall level of wealth, rather than its distribution within the country, is a defining factor.\\

Code Week exhibits an opposite trend compared to that reported in most of the previous work, where higher income has been associated with higher civic and social participation levels. Explanations for the inhibited participation in low-income and disadvantaged areas revolve around two main views: (a) the conditions necessary for fruitful social interaction, namely opportunities for personal acquaintance, equal status, and the ability to share common goals, come to lack, progressively eroding the social capital and depressing participation; (b) resources necessary for participation, either at the individual or community level, are unobtainable, impeding participation altogether. The counter-trend observed in this paper can be ascribed to the different nature of the Code Week campaign, which provides teachers from all over the world with the opportunity to join, connect and contribute to the initiative, regardless of their socio-economic status. Teachers in less developed areas who might feel isolated in their community, disappointed by the inadequate infrastructure, and disillusioned by their institutions, can find in Code Week a virtual community, transversal to all socio-economic levels and untied from political or other formal affiliations. They have the opportunity to form bonds and be part of a virtual network of like-minded individuals, which might not have been possible in their original community. Moreover, Code Week promotes the common goal of bringing coding and digital literacy to as many people as possible in an enjoyable and innovative way. In this shared vision, participating teachers are at the forefront of this digital dissemination. By populating the event map, teachers become aware of being part of a massive international undertaking. Hence, it can be argued that Code Week fosters participation in less developed areas because it restores its social capital by enabling those conditions that otherwise come lacking: it offers opportunities to encounter like-minded people, guarantees that all participants are treated equally, and encourages a shared vision for its community.\\

Code Week is also free and accessible to all. It does not pose any prerequisite or requirement for participation other than being able to connect to the platform to upload the activity. Hence, a lack of resources in disadvantaged areas does not deter participation because no resources are required. On the contrary, Code Week provides resources to its participants in terms of networking, mentoring, professional development and upskilling opportunities, original and engaging teaching material, challenges and awards, recognition, reputation and self-esteem from being part of an international transformative initiative. In this scenario, it could be argued that participation is not affected by the availability of resources necessary to participate but by the perceived value assigned to the resources gained. Assuming that the teacher is aware of the initiative, she/he will be likely to engage in the program if the perceived benefits outweigh the effort required. This generally consists of the time and energy to plan and conduct the activity, obtain permission from the school administrator (if needed), and upload the event details into the system. As the effort is generally limited, the decisive factor becomes the perceived benefit, which is likely to be strongly affected by the country's socio-economic status. In most developed countries, teachers and pupils have had full access to computers and broadband internet for the past 20 years. Being more exposed to technology, they are inherently more likely to become digitally competent. Computing education built into the curriculum and professional development programs promoted by the country's ministry of education provide a formal framework for learning and teaching computational thinking and digital skills. The country's stable social infrastructure can offer teachers plenty of networking and informal professional development opportunities. In this setting, Code Week does not always rise above the many other activities available to the teachers, who might choose not to participate. On the contrary, less developed countries have gained widespread access to modern technologies only in the last decade, with many rural areas still without internet or computer access. Computing education is either not contemplated or is only now being integrated into the curriculum. However, systematic professional development efforts are not yet in place, resulting in a severe skill gap. Teachers in less developed regions understand the importance of including computational thinking activities in their teaching and exposing students to these concepts from an early age but face the lack of resources, technology and opportunities for networking and upskilling in their community. They are aware of the importance of their role in society, but they feel under-resourced and under-appreciated, so they look for redemption opportunities. In this scenario, Code Week represents a unique opportunity that compensates for institutional shortcomings. Overall, Code Week appears to be more prevalent where it is most needed and can be most beneficial.\\


While participation varies significantly among EU countries, it appears to be relatively homogeneous within a country, especially in less developed ones, where event incidence is consistently high regardless of income. A country-wide participation trend is consistent with previous studies which identify factors affecting social and civic engagement at the society level. In other words, citizens of different areas of a country have a similar tendency to join the Code Week because the main affecting factors exist at the country level. Candidate socio-economic factors are the overall infrastructure level, access to technology, the quality of government (and the faith citizens have in their formal institutions), the human capital, and the degree of social cohesion. Code Week-related aspects could be the ambassador's effort to promote the vision, the number and distribution of leading teachers, the Ministry of Education's support of the initiative, the media coverage, and the competing or collaborative relationship with other coding dissemination initiatives. Education-related factors to consider are the teachers' level of expertise, the availability of continuous professional development, and the degree of integration of computing education in the compulsory curriculum. The latter is likely to affect Code Week participation directly and would be interesting to investigate. However, it is also difficult to quantify, depending on a broad set of interacting variables:  whether mandatory or optional, since what age, how many hours a week, as a separate subject or cross-curricular theme. The level of interest and action toward including CT in the curriculum varies significantly across Europe \cite{bocconi2022reviewing}: countries not planning to teach CT, finalising an updated framework, running a pilot program, integrating CT only recently, teaching CT for several years. Moreover, the time necessary to successfully deliver the updated framework after its approval is likely affected by the country's infrastructure and available resources. \\

The only significant within-country trend is observed in Italy, where participation is more prominent in less developed regions. Unlike most European countries, Italy is vastly heterogeneous from a political, economic and social point of view. The southern regions (denoted "Mezzogiorno") report an income per capita of less than two-thirds of that in the rest of the country. Differences in social structures affect the country's social integrity, with horizontal structures more common in the North and hierarchical forms more frequent in the South \cite{helliwell1995economic}. Civic community and citizen involvement tend to be more prominent in the North, where the government is also more efficient \cite{leonardi2001making}. The overall development gap between the two halves is likely to affect Code Week participation, but exact underlying mechanisms need to be further investigated.



\section{Conclusion and Future Work}

The study suggests that the counter-trend observed in Code Week compared to most other forms of civic and social participation can find an explanation by the peculiar features of the program: (a) it builds a social network of collaborations among like-minded people, (b) it does not require but offers resources to its participants, (c) the perceived added value of such resources is inversely proportional to their availability. Other types of social and civic engagements with comparable characteristics, e.g. Erasmus and eTwinning programs, or opposite features (e.g. require an initial investment, do not offer an immediate return, do not entail social interactions) should be compared to Code Week participation to validate this view further.\\

There is also an evident correspondence between the development classification of the EU Regional Fund and Code Week participation. Investigating a possible causal relationship between the two is beyond the scope of the work but deserves further consideration. The EU regional Fund allocates the most funding to less developed regions. Local organisations apply for funding by presenting a project in response to a regional call, and a significant portion is financed. These regions could be more eager to join Code Week because (a) they have an appreciation for the EU, having received funds to develop the region, or (b) have acquired project-managing skills, having already developed project proposals in a different context. Regression discontinuity on the boundary of regions with different levels of development would help clarify how the Regional Fund affects participation. If there exists a causal relationship, a survey-based approach should be used to identify the main underlying factor(s).\\

This study addressed the impact of Code Week in terms of its ability to reach and engage people, particularly those who would most benefit from this initiative. From this point of view, Code Week is undoubtedly a success story. The campaign's impact intended as computing skills acquired by participants would be more challenging to measure, mainly due to the lack of participants' feedback on the activities. However, that is a rather narrow and superficial definition of effectiveness, which does not reflect the campaign's vision, and for which more suitable tools have been developed, such as SELFIE \cite{costa2021capturing}. Code Week's global vision entails raising awareness about the importance of digital skills and the potential of modern technologies, challenging stereotypes and defying prevailing norms, fostering creativity and teamwork, and promoting coding skills for all ages. Participating in Code Week means taking an active stand in achieving these goals, hence is a more inclusive and appropriate measure of the effectiveness of this campaign.

\begin{table}
\caption{Penetration Statistics}
\label{tab:pen_results}
\begin{tabular}{c c c c c c c c c c c}
         ISO & events & creators & locations & creator & location & events & creators & locations & total & average \\ 
        ~ & number & number & number & relevance & relevance & incidence & incidence & incidence & coverage & coverage \\ \hline
        MT & 1294 & 420 & 273 & 3.1 & 4.7 & 262.2 & 93.2 & 60.6 & 1.00 & 0.94 \\ 
        EE & 2663 & 759 & 804 & 3.5 & 3.3 & 201.0 & 57.7 & 61.1 & 1.00 & 1.00 \\ 
        PL & 64654 & 13241 & 14314 & 4.9 & 4.5 & 170.3 & 34.9 & 37.7 & 1.00 & 0.98 \\ 
        IT & 94303 & 26275 & 17794 & 3.6 & 5.3 & 157.7 & 43.3 & 29.3 & 1.00 & 0.95 \\ 
        TR & 75041 & 30853 & 17578 & 2.4 & 4.3 & 91.5 & 39.2 & 22.3 & 1.00 & 0.75 \\ 
        HR & 3671 & 1132 & 1214 & 3.2 & 3.0 & 90.1 & 27.0 & 29.0 & 1.00 & 0.89 \\ 
        AL & 2240 & 479 & 460 & 4.7 & 4.9 & 78.3 & 16.7 & 16.0 & 1.00 & 0.57 \\ 
        LT & 2098 & 763 & 551 & 2.7 & 3.8 & 75.1 & 26.4 & 19.1 & 1.00 & 0.89 \\ 
        LU & 388 & 37 & 84 & 10.5 & 4.6 & 63.2 & 6.4 & 14.6 & 1.00 & 1.00 \\ 
        HU & 5891 & 877 & 1119 & 6.7 & 5.3 & 60.3 & 8.9 & 11.4 & 1.00 & 0.90 \\ 
        EL & 5555 & 2339 & 2427 & 2.4 & 2.3 & 51.8 & 21.7 & 22.5 & 1.00 & 0.82 \\ 
        MK & 1060 & 419 & 424 & 2.5 & 2.5 & 51.0 & 20.2 & 20.5 & 1.00 & 0.80 \\
        ME & 298 & 151 & 139 & 2.0 & 2.1 & 47.9 & 24.3 & 22.3 & 1.00 & 0.62 \\ 
        SI & 811 & 230 & 297 & 3.5 & 2.7 & 39.0 & 11.1 & 14.4 & 1.00 & 0.77 \\
        BG & 2544 & 792 & 927 & 3.2 & 2.7 & 36.3 & 11.1 & 13.0 & 1.00 & 0.83 \\
        RO & 6162 & 2375 & 2178 & 2.6 & 2.8 & 31.7 & 12.0 & 11.0 & 1.00 & 0.79 \\
        IE & 1529 & 487 & 816 & 3.1 & 1.9 & 31.2 & 10.3 & 17.3 & 1.00 & 1.00 \\
        RS & 2121 & 694 & 977 & 3.1 & 2.2 & 30.5 & 9.8 & 13.8 & 1.00 & 0.91 \\ 
        AT & 2442 & 595 & 635 & 4.1 & 3.8 & 27.6 & 6.8 & 7.3 & 1.00 & 0.42 \\
        CY & 241 & 86 & 112 & 2.8 & 2.2 & 27.5 & 10.1 & 13.2 & 1.00 & 1.00 \\
        LV & 527 & 222 & 235 & 2.4 & 2.2 & 27.4 & 11.3 & 11.9 & 1.00 & 0.77 \\ 
        PT & 2317 & 1021 & 1012 & 2.3 & 2.3 & 22.5 & 9.9 & 9.8 & 1.00 & 0.76 \\
        SK & 944 & 293 & 399 & 3.2 & 2.4 & 17.3 & 5.4 & 7.4 & 1.00 & 0.94 \\ 
        IS & 53 & 22 & 34 & 2.4 & 1.6 & 14.8 & 6.6 & 10.2 & 1.00 & 0.75 \\ 
        ES & 6270 & 2326 & 3000 & 2.7 & 2.1 & 13.4 & 5.0 & 6.5 & 1.00 & 0.74 \\ 
        SE & 1125 & 436 & 509 & 2.6 & 2.2 & 11.0 & 4.4 & 5.2 & 1.00 & 0.65 \\ 
        CZ & 950 & 304 & 391 & 3.1 & 2.4 & 8.9 & 2.9 & 3.7 & 1.00 & 0.84 \\
        BE & 1009 & 258 & 536 & 3.9 & 1.9 & 8.8 & 2.3 & 4.7 & 0.77 & 0.46 \\ 
        FI & 376 & 145 & 231 & 2.6 & 1.6 & 6.8 & 2.6 & 4.2 & 0.81 & 0.44 \\ 
        NL & 1019 & 158 & 307 & 6.4 & 3.3 & 5.9 & 0.9 & 1.8 & 0.85 & 0.38 \\
        DE & 4113 & 734 & 1262 & 5.6 & 3.3 & 5.0 & 0.9 & 1.5 & 0.55 & 0.18 \\ 
        CH & 415 & 113 & 118 & 3.7 & 3.5 & 4.9 & 1.4 & 1.4 & 0.70 & 0.28 \\
        FR & 2393 & 475 & 894 & 5.0 & 2.7 & 3.6 & 0.7 & 1.3 & 0.87 & 0.42 \\
        UK & 2245 & 240 & 448 & 9.4 & 5.0 & 3.4 & 0.4 & 0.7 & 0.66 & 0.23 \\
        DK & 108 & 28 & 80 & 3.9 & 1.4 & 1.9 & 0.5 & 1.4 & 0.82 & 0.44 \\
        NO & 78 & 40 & 51 & 2.0 & 1.5 & 1.5 & 0.8 & 1.0 & 0.77 & 0.31 \\ \hline
        mean & 8304 & 2495 & 2018 & 3.8 & 3.0 & 49.5 & 15.2 & 14.7 & 0.94 & 0.71 \\
        std err & 3515 & 1102 & 733 & 0.3 & 0.2 & 9.8 & 3.1 & 2.4 & 0.02 & 0.04 \\ 
\end{tabular}
\end{table}
    
\begin{table}
\caption{Demographic Statistics}
\label{tab:dem_results}
\begin{tabular}{c c c c c c c c c c c c c}
        ISO & \%Filled & Event & Part. & Female & \multicolumn{3}{c}{Audience Prevalence} & \multicolumn{3}{c}{Female Participation} & \multicolumn{2}{c}{Organiser Prevalence} \\ 
        ~ & Reports & Size & Age & Particip. & 4-14 & 14-24 & 24+ & 4-14 & 14-24 & 24+ & School & Business \\ \hline
        MT & 44.7 & 45 & 9.0 & 50.2 & 61.1 & 27.0 & 12.0 & 51.0 & 46.5 & 68.5 & 72.0 & 16.8 \\ 
        EE & 59.6 & 32 & 10.6 & 42.6 & 82.0 & 10.1 & 8.0 & 41.3 & 43.0 & 64.3 & 65.9 & 1.3 \\ 
        PL & 61.2 & 31 & 9.5 & 38.5 & 89.6 & 7.4 & 3.1 & 39.0 & 39.4 & 71.6 & 73.0 & 0.9 \\ 
        IT & 74.8 & 36 & 10.4 & 46.2 & 72.9 & 25.1 & 1.9 & 45.8 & 41.8 & 66.5 & 97.2 & 1.3 \\ 
        TR & 69.5 & 68 & 10.3 & 53.3 & 85.8 & 12.5 & 1.6 & 52.9 & 54.6 & 62.6 & 98.2 & 0.1 \\ 
        HR & 75.6 & 38 & 12.4 & 47.7 & 71.4 & 25.9 & 2.5 & 47.5 & 46.8 & 70.0 & 94.0 & 0.4 \\ 
        AL & 82.0 & 62 & 14.0 & 54.1 & 52.9 & 37.4 & 9.8 & 50.9 & 58.7 & 69.8 & 96.6 & 0.5 \\ 
        LT & 79.6 & 46 & 11.2 & 45.4 & 60.3 & 34.7 & 5.1 & 45.5 & 42.1 & 70.0 & 94.2 & 0.1 \\ 
        LU & 9.0 & 57 & 18.7 & 47.2 & 70.2 & 17.7 & 12.0 & 47.6 & 49.0 & 37.0 & 63.6 & 0.0 \\ 
        HU & 69.5 & 40 & 13.9 & 41.7 & 60.2 & 34.8 & 4.9 & 40.9 & 36.9 & 65.2 & 85.6 & 0.5 \\ 
        EL & 74.5 & 61 & 10.1 & 48.9 & 74.9 & 20.5 & 4.6 & 48.4 & 50.0 & 67.8 & 95.3 & 1.1 \\ 
        MK & 84.0 & 36 & 11.7 & 45.8 & 85.4 & 11.9 & 2.7 & 45.4 & 47.4 & 50.4 & 89.8 & 0.2 \\ 
        ME & 78.2 & 26 & 12.0 & 49.7 & 81.1 & 17.0 & 1.9 & 48.0 & 53.2 & 81.1 & 97.3 & 0.3 \\ 
        SI & 57.1 & 54 & 12.1 & 44.4 & 57.4 & 27.2 & 15.4 & 44.6 & 43.1 & 58.6 & 71.6 & 0.7 \\ 
        BG & 54.4 & 47 & 13.5 & 44.5 & 41.6 & 45.8 & 12.5 & 46.2 & 39.4 & 70.5 & 75.1 & 1.1 \\ 
        RO & 80.0 & 38 & 12.1 & 44.8 & 69.1 & 25.1 & 5.7 & 43.4 & 49.1 & 54.1 & 85.6 & 0.5 \\ 
        IE & 20.3 & 80 & 12.0 & 50.3 & 49.8 & 44.3 & 5.8 & 51.5 & 43.7 & 58.5 & 76.4 & 0.0 \\ 
        RS & 79.6 & 66 & 13.1 & 48.1 & 64.9 & 25.8 & 9.2 & 48.8 & 44.7 & 60.8 & 83.1 & 0.7 \\ 
        AT & 4.8 & 68 & 15.7 & 42.0 & 23.2 & 45.4 & 31.3 & 47.1 & 39.4 & 24.6 & 93.5 & 1.9 \\ 
        CY & 67.6 & 80 & 10.7 & 51.4 & 70.2 & 18.8 & 11.1 & 51.9 & 41.8 & 57.4 & 89.2 & 2.1 \\ 
        LV & 54.6 & 50 & 13.1 & 49.0 & 75.3 & 19.7 & 5.0 & 48.9 & 46.4 & 64.0 & 84.6 & 0.6 \\ 
        PT & 68.5 & 71 & 12.7 & 47.5 & 72.1 & 23.6 & 4.3 & 49.9 & 37.5 & 55.1 & 88.4 & 1.6 \\ 
        SK & 51.7 & 55 & 13.5 & 43.3 & 48.0 & 34.5 & 17.5 & 44.8 & 41.4 & 43.9 & 64.8 & 0.2 \\ 
        IS & 13.2 & 75 & 11.0 & 52.9 & 71.5 & 16.5 & 12.1 & 52.9 & 0.0 & 0.0 & 100.0 & 0.0 \\
        ES & 46.2 & 66 & 13.1 & 48.5 & 42.3 & 41.4 & 16.3 & 50.1 & 39.7 & 55.6 & 75.8 & 7.9 \\ 
        SE & 20.7 & 70 & 11.5 & 46.6 & 66.8 & 20.4 & 12.8 & 48.1 & 30.1 & 66.4 & 76.2 & 14.8 \\ 
        CZ & 56.7 & 50 & 13.8 & 46.8 & 56.6 & 22.1 & 21.3 & 44.5 & 49.4 & 65.0 & 56.9 & 6.2 \\ 
        BE & 30.3 & 79 & 17.3 & 53.0 & 39.9 & 41.2 & 18.8 & 48.9 & 59.6 & 58.6 & 42.1 & 7.9 \\ 
        FI & 32.2 & 51 & 14.7 & 46.5 & 54.5 & 23.1 & 22.3 & 46.6 & 41.6 & 49.3 & 71.5 & 0.8 \\
        NL & 51.8 & 32 & 11.5 & 47.4 & 56.9 & 29.6 & 13.5 & 48.9 & 45.2 & 46.9 & 65.0 & 17.9 \\ 
        DE & 23.8 & 35 & 14.8 & 42.4 & 29.8 & 48.4 & 21.9 & 40.3 & 44.6 & 43.0 & 20.7 & 21.0 \\ 
        CH & 19.5 & 33 & 11.1 & 47.5 & 31.3 & 39.5 & 29.2 & 46.7 & 48.9 & 73.7 & 26.6 & 56.2 \\ 
        FR & 16.6 & 44 & 13.4 & 44.6 & 32.6 & 39.7 & 27.7 & 45.3 & 42.7 & 40.7 & 27.1 & 54.0 \\ 
        UK & 8.1 & 98 & 11.1 & 47.9 & 30.9 & 41.7 & 27.3 & 49.7 & 42.0 & 58.8 & 11.0 & 82.4 \\ 
        DK & 13.9 & 18 & 18.8 & 35.3 & 53.1 & 21.9 & 25.0 & 22.2 & 33.9 & 54.7 & 37.0 & 7.4 \\
        NO & 17.9 & 44 & 10.3 & 59.9 & 56.1 & 34.1 & 9.9 & 59.5 & 0.0 & 78.0 & 68.0 & 0.0 \\ \hline
        mean & 48.7 & 52 & 12.6 & 47.1 & 59.5 & 28.1 & 12.4 & 46.8 & 44.5 & 59.5 & 72.6 & 9.7 \\ 
        std err & 4.2 & 3 & 0.4 & 0.7 & 2.8 & 1.8 & 1.4 & 0.9 & 1.1 & 2.0 & 3.9 & 3.3 \\ 
\end{tabular}
\end{table}
   
\begin{table}
\caption{Retention Statistics}
\label{tab:ret_results}
\begin{tabular}{c c c c c c c c c c c}
        ISO & Creator & Location & Creators & Locations & \multicolumn{3}{c}{Creators per \#editions} & \multicolumn{3}{c}{Locations per \#editions} \\ 
        ~ & Recruiting & Recruiting & Frequency & Frequency & 1  & 2  & 2+ & 1 & 2 & 2+ \\ \hline
        MT & 78 & 73 & 1.27 & 1.39 & 81.0 & 13.6 & 5.5 & 74.4 & 16.5 & 9.2 \\ 
        EE & 74 & 73 & 1.42 & 1.39 & 73.3 & 17.9 & 8.9 & 73.9 & 17.3 & 8.7 \\ 
        PL & 72 & 76 & 1.46 & 1.36 & 71.0 & 18.3 & 10.6 & 75.8 & 16.3 & 7.8 \\
        IT & 73 & 72 & 1.43 & 1.43 & 74.3 & 15.6 & 10.1 & 72.3 & 17.7 & 10.0 \\
        TR & 82 & 84 & 1.29 & 1.31 & 78.9 & 15.1 & 5.9 & 76.3 & 17.1 & 6.6 \\ 
        HR & 61 & 65 & 1.66 & 1.50 & 64.2 & 20.5 & 15.2 & 71.4 & 16.3 & 12.3 \\ 
        AL & 77 & 85 & 1.34 & 1.27 & 77.5 & 15.2 & 7.3 & 81.3 & 12.8 & 5.8 \\ 
        LT & 65 & 68 & 1.60 & 1.42 & 64.4 & 21.0 & 14.7 & 72.2 & 19.4 & 8.4 \\ 
        LU & 58 & 73 & 1.46 & 1.21 & 78.4 & 8.1 & 13.5 & 88.1 & 4.8 & 7.2 \\ 
        HU & 67 & 70 & 1.57 & 1.42 & 68.8 & 16.1 & 15.2 & 75.5 & 13.9 & 10.6 \\ 
        EL & 61 & 75 & 1.46 & 1.28 & 72.6 & 17.2 & 10.2 & 79.9 & 14.0 & 6.1 \\
        MK & 66 & 80 & 1.51 & 1.25 & 72.1 & 15.0 & 12.9 & 80.7 & 14.2 & 5.2 \\ 
        ME & 73 & 79 & 1.15 & 1.12 & 86.1 & 12.6 & 1.3 & 88.5 & 10.8 & 0.7 \\
        SI & 74 & 73 & 1.34 & 1.37 & 77.4 & 13.9 & 8.6 & 74.7 & 16.5 & 8.7 \\
        BG & 68 & 76 & 1.46 & 1.29 & 71.5 & 17.2 & 11.3 & 80.0 & 13.8 & 6.1 \\
        RO & 77 & 82 & 1.42 & 1.27 & 72.6 & 17.3 & 10.3 & 78.7 & 16.4 & 4.9 \\
        IE & 75 & 84 & 1.27 & 1.12 & 81.7 & 11.9 & 6.3 & 90.4 & 7.8 & 1.7 \\ 
        RS & 59 & 71 & 1.61 & 1.33 & 68.0 & 17.9 & 14.1 & 77.4 & 15.8 & 6.8 \\ 
        AT & 77 & 78 & 1.22 & 1.23 & 80.2 & 18.3 & 1.6 & 79.7 & 18.4 & 1.9 \\ 
        CY & 61 & 77 & 1.57 & 1.21 & 68.6 & 17.4 & 14.0 & 85.7 & 10.7 & 3.6 \\ 
        LV & 74 & 83 & 1.36 & 1.21 & 72.5 & 20.7 & 6.9 & 84.7 & 10.6 & 4.7 \\ 
        PT & 74 & 82 & 1.34 & 1.22 & 77.4 & 15.4 & 7.3 & 83.0 & 13.4 & 3.6 \\ 
        SK & 55 & 66 & 1.50 & 1.34 & 73.7 & 13.7 & 12.6 & 80.2 & 11.8 & 8.1 \\ 
        IS & 76 & 94 & 1.18 & 1.06 & 86.4 & 9.1 & 4.5 & 94.1 & 5.9 & 0.0 \\ 
        ES & 70 & 74 & 1.41 & 1.33 & 75.3 & 15.0 & 9.7 & 78.0 & 14.6 & 7.6 \\ 
        SE & 76 & 77 & 1.20 & 1.18 & 83.9 & 13.1 & 3.0 & 86.6 & 10.4 & 3.0 \\ 
        CZ & 66 & 72 & 1.57 & 1.42 & 68.8 & 15.8 & 15.5 & 73.1 & 16.6 & 10.3 \\ 
        BE & 70 & 76 & 1.37 & 1.27 & 75.2 & 15.9 & 8.9 & 81.5 & 12.7 & 5.8 \\ 
        FI & 72 & 85 & 1.20 & 1.12 & 85.5 & 10.3 & 4.2 & 90.0 & 8.7 & 1.3 \\ 
        NL & 81 & 87 & 1.20 & 1.12 & 86.1 & 8.9 & 5.0 & 89.6 & 9.1 & 1.3 \\ 
        DE & 81 & 82 & 1.28 & 1.27 & 78.9 & 15.5 & 5.6 & 79.7 & 14.9 & 5.5 \\ 
        CH & 85 & 74 & 1.15 & 1.27 & 88.5 & 8.8 & 2.7 & 78.0 & 16.9 & 5.1 \\ 
        FR & 76 & 81 & 1.25 & 1.20 & 83.8 & 10.1 & 6.0 & 86.8 & 8.4 & 4.8 \\ 
        UK & 85 & 83 & 1.13 & 1.21 & 90.4 & 7.5 & 2.0 & 85.7 & 8.3 & 6.0 \\ 
        DK & 75 & 91 & 1.25 & 1.08 & 89.3 & 7.1 & 3.6 & 92.5 & 7.5 & 0.0 \\ 
        NO & 81 & 91 & 1.15 & 1.08 & 87.5 & 10.0 & 2.5 & 92.2 & 7.8 & 0.0 \\ \hline
        mean & 72.1 & 78.1 & 1.36 & 1.27 & 77.4 & 14.4 & 8.3 & 81.5 & 13.0 & 5.5 \\
        std err & 1.2 & 1.1 & 0.02 & 0.02 & 1.2 & 0.6 & 0.7 & 1.1 & 0.6 & 0.5 \\ 
\end{tabular}
\end{table}   
    
\begin{table}
\caption{Spatial and Trend Statistics}
\label{tab:spa_results}
\begin{tabular}{c c c c c c c | c c c}
        ISO & Neighbours & Neighbours & \multicolumn{2}{c}{Creators} & \multicolumn{2}{c}{New Creators} & \multicolumn{3}{c}{Trend} \\ 
        ~ & per Creator & per Location & previous loc. & near loc. & previous loc. & near loc. & 2014-21 & 2014-19 & diff. \\ \hline
        MT & 25.2 & 13.6 & 92.7 & 5.4 & 91.5 & 5.7 & 3.18 & 9.19 & -6.01 \\ 
        EE & 3.6 & 2.9 & 71.9 & 11.8 & 64.7 & 13.9 & 1.57 & 7.18 & -5.61 \\ 
        PL & 3.9 & 3.2 & 72.8 & 13.0 & 64.0 & 16.7 & 2.35 & 9.33 & -6.98 \\ 
        IT & 8.2 & 4.8 & 75.4 & 11.2 & 71.4 & 13.2 & 0.03 & 2.31 & -2.28 \\ 
        TR & 9.4 & 4.5 & 61.3 & 14.0 & 58.3 & 13.5 & 1.99 & 5.93 & -3.94 \\ 
        HR & 3.3 & 3.0 & 79.7 & 9.7 & 67.0 & 13.6 & 1.13 & 4.41 & -3.28 \\ 
        AL & 4.3 & 2.7 & 61.2 & 8.7 & 59.5 & 6.4 & 1.25 & 5.33 & -4.08 \\ 
        LT & 5.3 & 3.2 & 73.8 & 10.1 & 63.7 & 12.7 & 0.75 & 3.45 & -2.70 \\ 
        LU & 2.2 & 2.3 & 68.3 & 18.7 & 62.5 & 25.5 & 1.37 & 2.66 & -1.29 \\
        HU & 3.1 & 2.7 & 76.8 & 7.2 & 62.1 & 11.2 & 0.49 & 2.10 & -1.61 \\
        EL & 4.3 & 3.5 & 79.3 & 9.4 & 71.4 & 11.9 & 0.58 & 1.62 & -1.04 \\
        MK & 4.7 & 3.7 & 68.8 & 9.0 & 66.4 & 9.3 & 0.55 & 2.65 & -2.10 \\ 
        ME & 2.8 & 2.2 & 41.3 & 17.7 & 36.5 & 22.0 & 0.70 & 4.17 & -3.47 \\
        SI & 2.6 & 2.3 & 68.4 & 9.9 & 54.4 & 16.7 & 0.08 & 1.15 & -1.07 \\ 
        BG & 4.8 & 3.7 & 71.6 & 12.8 & 70.7 & 11.9 & 0.35 & 1.87 & -1.52 \\ 
        RO & 4.5 & 3.1 & 61.4 & 9.6 & 54.3 & 11.1 & 0.60 & 2.02 & -1.42 \\ 
        IE & 2.3 & 2.3 & 68.7 & 18.6 & 64.5 & 20.8 & -0.78 & -0.28 & -0.50 \\ 
        RS & 3.2 & 3.3 & 79.6 & 7.9 & 73.9 & 10.5 & -0.11 & 0.64 & -0.75 \\ 
        AT & 1.8 & 1.7 & 68.3 & 11.9 & 55.5 & 18.5 & 0.76 & 0.68 & 0.08 \\ 
        CY & 2.4 & 2.2 & 68.9 & 16.2 & 55.0 & 19.7 & 0.19 & 1.27 & -1.08 \\ 
        LV & 2.2 & 1.8 & 44.2 & 21.5 & 32.6 & 28.8 & 0.41 & 1.26 & -0.85 \\ 
        PT & 3.5 & 2.8 & 64.9 & 17.8 & 58.7 & 19.7 & 0.29 & 1.13 & -0.84 \\ 
        SK & 2.4 & 2.5 & 79.8 & 9.7 & 62.2 & 15.8 & -0.06 & 0.40 & -0.46 \\ 
        IS & 1.7 & 1.9 & 59.8 & 20.2 & 41.7 & 28.3 & -0.53 & -0.37 & -0.16 \\ 
        ES & 2.8 & 2.8 & 74.5 & 11.0 & 63.4 & 14.4 & 0.00 & 0.37 & -0.37 \\ 
        SE & 2.4 & 2.2 & 68.8 & 14.0 & 59.4 & 14.1 & -0.08 & 0.24 & -0.32 \\ 
        CZ & 2.2 & 2.0 & 63.0 & 12.4 & 49.7 & 16.3 & 0.07 & 0.35 & -0.28 \\ 
        BE & 2.5 & 2.3 & 70.2 & 14.0 & 71.9 & 15.2 & 0.00 & 0.25 & -0.25 \\ 
        FI & 1.8 & 1.8 & 55.9 & 15.8 & 42.1 & 22.6 & -0.05 & 0.11 & -0.16 \\ 
        NL & 2.0 & 1.9 & 64.4 & 16.3 & 51.3 & 28.9 & 0.10 & 0.16 & -0.06 \\ 
        DE & 2.4 & 2.3 & 67.7 & 14.0 & 55.7 & 16.2 & 0.06 & 0.23 & -0.17 \\
        CH & 2.7 & 2.1 & 75.1 & 12.5 & 66.3 & 16.3 & -0.02 & 0.08 & -0.10 \\ 
        FR & 2.1 & 2.3 & 69.6 & 12.8 & 52.2 & 12.5 & -0.03 & 0.06 & -0.09 \\ 
        UK & 1.7 & 1.6 & 53.7 & 18.2 & 35.3 & 21.0 & -0.01 & 0.11 & -0.12 \\ 
        DK & 1.5 & 1.5 & 51.7 & 6.3 & 68.8 & 14.1 & -0.02 & -0.02 & 0.00 \\ \
        NO & 1.4 & 1.5 & 38.8 & 17.1 & 21.8 & 12.6 & -0.04 & -0.06 & 0.02 \\ \hline
        mean & 3.8 & 2.9 & 67.0 & 13.0 & 58.3 & 16.2 & 0.50 & 2.00 & -1.50 \\ 
        sem & 0.7 & 0.3 & 1.8 & 0.7 & 2.2 & 0.9 & 0.10 & 0.40 & 0.30 \\
\end{tabular}
\end{table}  

\bibliographystyle{ACM-Reference-Format}
\bibliography{codeweek}

\end{document}